\documentclass[journal, letterletter, final, 10pt]{IEEEtran}
\usepackage[dvips]{graphics}
\usepackage[dvips]{changebar}
\usepackage{amsmath,bm}
\usepackage{rotating}
\usepackage[strings]{underscore}
\usepackage{setspace}
\usepackage{times}
\usepackage{color}
\usepackage{soul}
\usepackage{amsmath} 
\usepackage{amssymb}
\usepackage[strings]{underscore}
\usepackage{graphics,color}
\usepackage{epstopdf}
\usepackage{graphicx}
\usepackage{multirow}
\usepackage{ifpdf}
\usepackage{tablefootnote}
\usepackage{algorithmicx}
\usepackage{algpseudocode}
\usepackage[font=footnotesize, skip=0pt]{caption}

\newcommand{\mb}[1]{{\mathbf #1}}

\begin{document}
	%------------------------------------------------------------------------------------------------------------------------------------------------
	\setcounter{page}{1}
	\newcounter{mytempeqncnt}
	\setlength{\textfloatsep}{-1pt}
	\setlength{\intextsep}{1pt}
	%------------------------------------------------------------------------------------------------------------------------------------------------
	\title{Channel Covariance Matrix Estimation via Dimension Reduction for Hybrid MIMO MmWave Communication Systems}

	%\squeezeup
	\author{Rui Hu, Jun Tong, Jiangtao Xi, Qinghua Guo and Yanguang Yu}
	%	\thanks{
		%	The authors are with the School of Electrical, Computer and Telecommunications Engineering, University of Wollongong, Wollongong, NSW 2522, Australia. Email:  rh546@uowmail.edu.au, \{jtong, jiangtao, qguo, yanguang\}@uow.edu.au. 
		%	\emph{Corresponding author: J. Tong}.} 
%	} 
	
	%\setlength{\droptitle}{-5em}
%	\date{}
	\maketitle
	
	%\thispagestyle{empty}
	%\pagestyle{plain}	
	%------------------------------------------------------------------------------------------------------------------------------------------------
	%\baselineskip=27pt
	%------------------------------------------------------------------------------------------------------------------------------------------------                                   
	
	\begin{abstract}
		Hybrid massive MIMO structures with lower hardware complexity and power consumption have been considered as a potential candidate for millimeter wave (mmWave) communications.
		Channel covariance information can be used for designing transmitter  precoders, receiver combiners, channel estimators, etc. However, 
		hybrid structures allow only a lower-dimensional signal to be observed, 
		which  
		adds difficulties for channel covariance matrix estimation.  In this paper, we formulate the channel covariance estimation as a structured low-rank matrix sensing problem via Kronecker product expansion and use   
		a low-complexity algorithm to solve this problem. 
		Numerical results with uniform linear arrays (ULA) and uniform squared planar arrays (USPA) are provided to demonstrate the effectiveness of our proposed method. 
	\end{abstract}
	
	\begin{IEEEkeywords}
		Millimeter wave communications, hybrid system, Kronecker product expansion, low-rank matrix recovery
	\end{IEEEkeywords}
	
	\maketitle

	\section{Introduction}
	Millimeter wave (mmWave) communications are promising for future-generation wireless communications for their advantages such as %extremely 
	large bandwidths, narrow beams, and secure transmissions \cite{SurveyI, SurveyII}.
	Large-scale multiple-input multiple-output (MIMO) 
	hybrid structures %, e.g., phase shifter- or switch-based, that 
	equipped with only a few %number of 
	RF chains have generated great interests %considerations 
	for mmWave systems due to their low complexity and near-optimal performance \cite{CMI,CMII}. To exploit the potential of %make the best advantage} of 
	large-scale MIMO hybrid systems, e.g., to achieve high data transmission rates, precoders and combiners must be carefully designed. They can be designed based on the instantaneous channel matrix \cite{PrecodingI, PrecodingII}, which may be estimated %is unknown and obtained 
	by using channel estimation techniques \cite{CMI, CMOMP}. However, the instantaneous channel can vary fast at mmWave frequencies and the precoder/combiner have to be redesigned once the instantaneous channel changes.

	Although the instantaneous mmWave channel can change very fast, the long-term channel statistics, e.g., the angular power spectrum, can be stationary for tens to hundreds of coherence blocks \cite{Joint}. 
	Recently, the channel covariance information has been utilized to design the analog precoders/combiners \cite{JSDM, Joint}, %. The \tj{resulting} analog precoders/combiners %do not need to be redesigned as long as the channel 
	which remain fixed when the covariance matrix is %long-term statistics are
	unchanged. 
	The effective digital system has a reduced dimensionality, which greatly reduces the cost for acquiring the instantaneous channel state information (CSI) and simplifies the optimization of the digital precoders and combines.
	To realize the designs in \cite{JSDM, Joint}, the channel covariance matrix should % information has to 
	be firstly estimated.  
	With large antenna arrays, the channel covariance matrix has a large dimensionality, which demands a large number of %training samples 
	observations to be used when traditional covariance matrix estimators are adopted. Meanwhile, the hybrid structure only allows a reduced number of observations to be acquired at the receiver, 
	which makes the channel covariance estimation task challenging. 
	%
	%Although the instantaneous channel can vary fast at mmWave frequencies, the long-term channel statistics, e.g., angular power spectrum, can be stationary for tens to hundreds of coherence blocks \cite{Joint}. Channel covariance information can be used for designing analog precoders/combiners \cite{Joint} or can be used 
	%for channel estimation \cite{Channel estimation for TDD/FDD}.
	%so that the dimension of channel to be seen at the baseband will be reduced to the same level as the number of RF chains \cite{Joint}; 
	%it also can be used for channel estimation, e.g., by using the minimum mean-squared error (MMSE) estimator \cite{Channel estimation for TDD/FDD}. 
	In order to address this challenge, \cite{CS} proposes several compressive sensing (CS) based channel covariance estimators, which
	%\tj{Their methods} 
	explore the relations between the angle of departure (AoD)/angle of arrival (AoA) and the channel covariance matrix. % to tackle the challenge. 
	Their methods need a dictionary for searching the AoD/AoA and the resulting performance improves when the resolution of the dictionary becomes higher. However, high-resolution dictionary yields high computational complexity. Moreover, these CS-based estimators require the number of paths in the channel to be known as a prior. In \cite{Channel estimation for TDD/FDD}, an analytical expression of the channel covariance matrix is derived and computed through the information obtained from one instantaneous channel realization, which can be estimated from low-dimensional observations. However, the analytical expression is given only for vector channels, %receivers with single antennas, 
	which may not be suitable for mmWave communications.

	In this paper, we investigate the mmWave channel covariance matrix estimation problem for hybrid mmWave communication systems that are equipped with uniform linear arrays (ULA) or uniform square planar arrays (USPA).  
	%We formulate the channel covariance estimation problem as a structured low-rank matrix sensing problem by exploiting Kronecker product expansion	\cite{Kronecker Covariance Sketching} and the structures of the ULA/USPA. 
	The main contributions are as follows: 
	\begin{enumerate}
		\item We show that the mmWave MIMO channel covariance matrix %is a summation of Kronecker products and it satisfies 
		follows a Kronecker product expansion model \cite{Kronecker Covariance Sketching}. 
		Following \cite{Kronecker Covariance Sketching, HeroI, HeroII}, 
		% Recently, 
		we show that this model can be % has been 
		used for reducing the effective dimension of the large-dimensional channel covariance matrices in mmWave MIMO systems.
		We further show that permutation can reduce the rank of the mmWave channel covariance matrix, which admits an expression of the  summation of vector outer products. We thus  
		formulate the channel covariance matrix estimation problem 
		as a low-rank matrix sensing problem.  
		
		\item Although the aforementioned low-rank matrix sensing problem has a lower size than the original problem, the complexity can still be high when the numbers of the transmitter/receiver antennas are large. %,problem size is still large. 
		In order to reduce the complexity, we further exploit %then explore 
		the structures of the ULA or USPA to reduce the dimensionality of the problem and formulate the problem as a \emph{structured} low-rank matrix sensing problem. %The dimensionality of the formulated problem is greatly reduced. 
We adapt the recently proposed generalized conditional gradient and alternating minimization (GCG-Alt) algorithm \cite{MCRui}, which has a low computational complexity, to find the solution. Numerical results with ULA and USPA suggest that our proposed estimator is effective in estimating the mmWave channel covariance matrix. 
		
	\end{enumerate}
	
	%The formulated problem has a reduced dimensionality. We  use the generalized conditional
	%gradient and alternating minimization (GCG-Alt) algorithm \cite{MCRui} to find the solution. %solve this problem. 
	%Our proposed estimator has lower complexity and better performance than the CS-based channel covariance estimator \cite{CS}.

	The rest of this paper is organized as follows. We introduce the spatial channel model and the hybrid system in Section II. In Section III, we formulate the channel covariance estimation problem as a structured low-rank matrix sensing problem and present the solution. We show the simulation results in Section IV and conclude the paper in Section V. 
	
	\emph{Notations}: Bold uppercase $\mb A$ denotes a matrix and bold lowercase $\mb a$ denotes a column vector. $\mb A^{\ast}$, $\mb A^T$, and $\mb A^H$ denote the conjugate, transpose, and conjugate transpose of matrix $\mb A$, respectively. $\mb a(i)$ denotes the $i$-th element of vector $\mb a$. $[\mb A]_{a:b,:}$ denotes the submatrix of $\mb A$ made of its $a$-th to $b$-th rows. % of $\mb A$ and all its columns. 
	$[\mb A]_{a:b,c:d}$ denotes the submatrix of $\mb A$ defined by its $a$-th to $b$-th rows and $c$-th to $d$-th columns. $\|\mb A\|_F$ and $\|\mb A\|_{\ast}$ are the Frobenius norm and the nuclear norm of $\mb A$. For $\mb A\in\mathbb{C}^{M\times N}$, $\text{vec}(\mb A)\in\mathbb{C}^{MN \times 1}$ is a column vector obtained through the vectorization of $\mb A$ and $\text{vec}^{-1}(\mb A)\in\mathbb{C}^{M\times N}$ is a matrix obtained by the inverse of vectorization.  For matrices $\mb A$ and $\mb B$, $\mb A\otimes \mb B$ denotes the Kronecker product of $\mb A$ and $\mb B$. $\mathcal{CN}(a,b^2)$ represents complex Gaussian distribution with mean $a$ and variance $b^2$. $\mathcal{U}(a,b)$ represents uniform distribution with support $[a,b]$.
	
	%
	%	\vspace{-1ex}
	\section{Spatial Channel Model}% And System}
	%In this paper, we 
	Consider %the downlink 
	point-to-point mmWave transmissions, where the transmitter has $N_t$ antennas and the receiver has $N_r$ antennas.  
	We assume 
	the following spatial channel \cite{mmWave Channel}:
	%\vspace{-1ex}
	\begin{equation}
	\label{mmWave H}
	\mb H=\frac{1}{\sqrt{L}}\displaystyle\sum_{k=1}^K\displaystyle\sum_{l=1}^Lg_{kl}\mb a_{r}(\phi^{r}_{kl}, \theta^{r}_{kl})\mb a_{t}^H(\phi^{t}_{kl},\theta^t_{kl})\in\mathbb{C}^{N_r\times N_t},
	\end{equation}
	%where $K\sim\text{max}\{\text{Poisson}(\lambda),1\}$ with $\lambda$ as the mean of the Possion distribution,
	where %$\mb H\in\mathbb{C}^{N_r\times N_t}$, 
	$K$ %(e.g., $K=1,2$) 
	is the number of clusters, and $L$ %(e.g., $L=30$) 
	is the number of rays within each cluster. As reported in \cite{mmWave Channel}, the number of clusters is often small, e.g., $K=1,2$, but the number of rays inside each cluster can be large, e.g., $L=30$.
	$\mb a_r(\phi^r_{kl},\theta^r_{kl})$ and $\mb a_t(\phi^t_{kl},\theta^t_{kl})$ are the array response vectors at the receiver and transmitter, respectively, where $\phi^r_{kl},\theta^r_{kl},\phi^t_{kl}$, and $\theta^t_{kl}$ are the azimuth AoA, elevation AoA, azimuth AoD, and elevation AoD on the $l$-th ray of the $k$-th cluster, respectively. 
	These angles can be characterized by cluster center angles and angular spreads: Each cluster covers a range of angles and the angular spread describes the span of each cluster. 
	%and ray angle shifts. Denote by $\phi^r_{k}, \theta^r_k, \phi^t_k$, and $\theta^t_k$ the azimuth AoA, the elevation AoA, the azimuth AoD, and the elevation AoD center angles of the $k$-th cluster, respectively. 
	%\dc{Let $\varphi^r_{kl}, \vartheta^r_{kl}, \varphi^t_{kl}$, and $\vartheta^t_{kl}$ be the azimuth AoA, the elevation AoA, the azimuth AoD, and the elevation AoD ray angle shifts 
	%	\textcolor{green}{[Can we simplify the modeling, e.g., avoid introducing these shifts if they are not used a lot?]} 
	%of the $l$-th ray away from the center angle of the $k$-th cluster. Then we have $\phi^r_{kl}=\phi^r_k-\varphi^r_{kl}, \theta^r_{kl}=\theta^r_k-\vartheta^r_{kl}, \phi^t_{kl}=\phi^t_k-\varphi^t_{kl}$, and $\theta^t_{kl}=\theta^t_k-\vartheta^t_{kl}$. 
	%This representation indicates that each cluster covers a range of angles and the angular spread describes the span of each cluster.} 
	The angular spread in the mmWave propagation environment is considered to be small \cite{ Channel estimation for TDD/FDD}. 
	Measurements of the angular spread taken in the urban area of New York City are presented in \cite{mmWave Channel} in terms of the root-mean-square (rms) of all the measurements. At the carrier frequency $f_c=28$ GHz, example angular spreads of $15.5^{\circ}, 6^{\circ}, 10.2^{\circ}$, and $0^{\circ}$ are reported for $\phi^r_{kl},\theta^r_{kl},\phi^t_{kl}$, and $\theta^t_{kl}$, %the azimuth AoA, the elevation AoA, the azimuth AoD, and the elevation AoD, 
	respectively. 
	The %complex 
	small-scale fading 
	coefficient %gain 
	$g_{kl}$ %on the $l$-th ray of the $k$-th cluster 
	is assumed complex Gaussian, i.e., %follows a complex Gaussian distribution, i.e., 
	$g_{kl}\sim\mathcal{CN}(0,\gamma^2_k)$, where $\gamma^2_{k}$ is the fraction power of the $k$-th cluster %and can be calculated using 
	\cite[Eq. (7)]{mmWave Channel}.
	%
	%The receiving and transmitting array response vectors are given by $\mb a_r(\phi^r_{kl})$ and $\mb a_t(\phi^t_{kl})$, respectively.
	%On the $l$-th ray of the $k$-th cluster, the AoA $\phi^r_{kl}=\phi^r_k-\varphi^r_{kl}$ and the AoD $\phi^t_{kl}=\phi^t_k-\varphi^t_{kl}$, where $\phi^r_{k}$ and $\phi^t_{k}$ are the cluster center angles seen at the receiver and transmitter, respectively, $\varphi^r_{kl}$ and $\varphi^t_{kl}$ are the $l$-th ray shifts of their respective clusters.
	%are the AoA and AoD on the $l$-th ray of the $k$-th cluster, respectively. 
	%The angles are characterized by cluster center angles and ray angle shifts, e.g., $\phi^r_{kl}=\phi^r_k-\varphi^r_{kl}$, where $\phi^r_{k}$ is the center angle of the $k$-th cluster be seen from the receiver side and $\varphi^r_{kl}$ is the angle shift of the $l$-th ray %away from the center angle of that cluster. 
	%Similarly, $\phi^t_{kl}=\phi^t_k-\varphi^t_{kl}$. 
		\begin{figure}
		\centering
		\label{Sys}
		\includegraphics[width=\columnwidth]{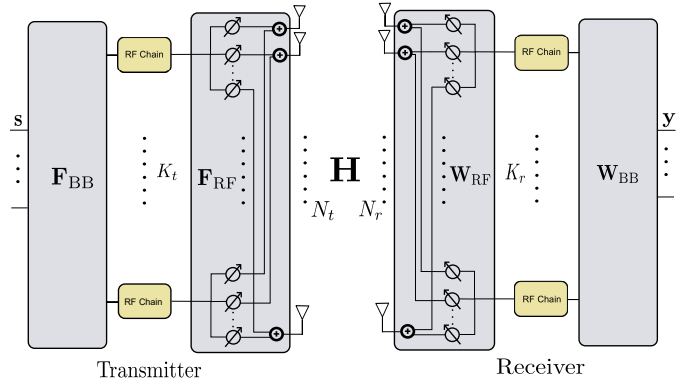}
						\vspace{0.1ex}
		\caption{The phase shifter-based hybrid transceiver. }
	\end{figure}
	
	As discussed in \cite{Joint}, though the small-scale fading gains $\{g_{kl}\}$ change fast, the AoDs/AoAs and $\gamma^2_k$ may remain stationary over tens to hundreds of coherence blocks. %By averaging small scale fading realizations, t
	Assume that $\{g_{kl}, \forall k,\forall l\}$ are mutually independent.
	The channel covariance matrix can then be modeled as 
	%	\vspace{-1ex}
	\begin{align}
	\label{R}\nonumber
	\mb R&\triangleq\mathbb{E}[\text{vec}(\mb H)\text{vec}^{H}(\mb H)]\\ 
	%&={ \frac{1}{ {L}}} \displaystyle\sum_{k=1}^K\gamma^2_k\displaystyle\sum_{l=1}^L(\mb a^{\ast}_{{t}_{kl}}\mb a^T_{{t}_{kl}})\otimes (\mb a_{{r}_{kl}}\mb a^H_{{r}_{kl}})\\
	&= { \frac{1}{ {L}}} \displaystyle\sum_{k=1}^K\gamma^2_k\displaystyle\sum_{l=1}^L{\widetilde{\mb T}^t_{{kl}}\otimes \widetilde{\mb T}^r_{{kl}}}\in\mathbb{C}^{N_rN_t\times N_rN_t},
	\end{align}
	where 
	%$\mb R\in\mathbb{C}^{N_rN_t\times N_rN_t}$,
	\begin{equation}
	\label{T_tkl}
	\widetilde{\mb T}^t_{{kl}} \triangleq \mb a^{\ast}_t(\phi^t_{kl},\theta^t_{kl})\mb a^T_t(\phi^t_{kl},\theta^t_{kl})\in\mathbb{C}^{N_t\times N_t},
	\end{equation}
	and
	\begin{equation}
	\label{T_rkl}
	\widetilde{\mb T}^r_{{kl}}\triangleq\mb a_r(\phi^r_{kl},\theta^r_{kl})\mb a^H_r(\phi^r_{kl},\theta^r_{kl})\in\mathbb{C}^{N_r\times N_r}.
	\end{equation} 
	Note that expression (\ref{R}) is the same as the channel covariance expression in \cite{Joint} when $L=1$.
	%For notational simplicity, we assume the transmitter and receiver adopt uniform linear arrays (ULA).
	In the following, we first present our proposed covariance matrix estimation method for systems equipped with the ULA and then discuss its adaptation to %show its implementation for 
	systems that adopt the USPA. 
	
	% \dc{Note that the elevation angle does not exist in the ULA,} so $\mb a_t(\phi^t_{kl},\theta^t_{kl})$ becomes $\mb a_t(\phi^t_{kl})$ and $\mb a_r(\phi^r_{kl},\theta^r_{kl})$ becomes $\mb a_r(\phi^r_{kl})$. 

	For the ULA, the array responses  $\mb a_t(\phi^t_{kl},\theta^t_{kl})$ and $\mb a_r(\phi^r_{kl},\theta^r_{kl})$ are independent of the elevation angles. They can thus be abbreviated as $\mb a_t(\phi^t_{kl})$ and  $\mb a_r(\phi^r_{kl})$.
	For an $N_a$-element ULA %placed along the $y$ axis 
	with distance $d$ between adjacent antennas, the array response is

	\begin{equation} 
	\label{aMS}\nonumber
	\mb a(\phi_{kl})=\frac{1}{\sqrt{N_a}}[1, \mathrm{e}^{j \frac{2\pi}{\lambda_c }d\sin(\phi_{kl})},
	\cdots,\mathrm{e}^{j(N_a-1)\frac{2\pi}{\lambda_c}d\sin(\phi_{kl})} ]^T, 
	\end{equation} 
	where $\lambda_c$ is the carrier wavelength and $N_a=N_t$ or $N_r$ is the number of antennas at the transmitter or receiver. %For ULA, 
	Accordingly, $\widetilde{\mb T}^t_{{kl}}$ of (\ref{T_tkl}) and $\widetilde{\mb T}^r_{{kl}}$ of (\ref{T_rkl}) become
	\begin{equation}
	\label{T_tklULA}
	\widetilde{\mb T}^t_{{kl}}=\mb a^{\ast}_t(\phi^t_{kl})\mb a^T_t(\phi^t_{kl})
	\end{equation}
	and 
	\begin{equation}
	\label{T_rklULA}
	\widetilde{\mb T}^r_{{kl}}=\mb a_r(\phi^t_{kl})\mb a^H_r(\phi^t_{kl}),
	\end{equation}
	respectively, which are %and they are 
	Toeplitz-Hermitian. Since the Kronecker product of two Toeplitz-Hermitian matrices is block-Toeplitz-Hermitian \cite{KP}, the channel covariance matrix $\mb R$ defined in (\ref{R}) is block-Toeplitz-Hermitian. 
	%and the summation of block Toeplitz Hermitian matrices is still a block Toeplitz Hermitian matrix, 
	%
	%\subsection{Hybrid Transceivers}
	
	We next discuss the hybrid system. 
	We assume phase shifter-based hybrid transceivers \cite{MCRui} shown in Fig. 1, where the antennas and analog phase shifters at the transmitter or receiver are fully connected. Assume that there are $K_t\ll N_t$ radio frequency (RF) chains at the transmitter and $K_r\ll N_r$ RF chains at the receiver.
	For single-stream transmissions with one symbol $s$ transmitted, the received signal is written as 
	%\vspace{-1ex}
	\begin{equation}
	\mb y=\mb W^H\mb H\mb f s+\mb W^H\mb n,
	\end{equation}
	where $\mb W$ and $\mb f$ are the receiving processing matrix and transmitting processing vector, respectively, and $\mb n$ is the noise vector. Up to $K_r$ digital symbols can be observed at the receiver after each transmission.  
	In hybrid transceivers, we have $\mb W=\mb W_{\rm{RF}}\mb W_{\rm{BB}}$ and $\mb f=\mb F_{\rm{RF}}\mb f_{\rm{BB}}$, where $\mb W_{\rm{RF}}$ and $\mb F_{\rm{RF}}$ are the analog combiner and precoder, respectively, and $\mb W_{\rm{BB}}$ and $\mb f_{\rm{BB}}$ are the digital combiner and precoder, respectively. In addition, due to the constraints of the phase shifters in the RF combiner and precoder, the entries in $\mb W_{\rm{RF}}$ and $\mb F_{\rm{RF}}$ have constant modulus. 
	Note that using single-stream transmissions during the channel training avoids the interferences caused by transmitting multiple symbols simultaneously and this has been widely considered \cite{ CMI, CMII, CMOMP}.
	
	%We can see that w
	When $N_t$ and $N_r$ are large, the dimension of the channel covariance matrix $\mb R$ is large. In this case, estimating $\mb R$ can be difficult when %we 
	only %have 
	a small number of observations available, which is typical in %under 
	the hybrid system. 
	%Recently, Kronecker product expansion has been used for reducing the effective dimension of large-dimensional covariance matrices that satisfy the Kronecker product expansion model \cite{Kronecker Covariance Sketching, KPExp}. 
	From (\ref{R}), %we can see that 
	$\mb R$ follows %is a summation of Kronecker products and satisfies 
	the Kronecker product expansion model \cite{Kronecker Covariance Sketching}. In the following, we explore this property and %the Kronecker and  
	the block-Toeplitz-Hermitian structure of $\mb R$ to reduce the dimensionality of the problem of estimating $\mb R$, and formulate the channel covariance matrix estimation problem as a structured low-rank matrix sensing problem. 
	
	%
	%\vspace{-10ex}
	\section{Structured Low-Rank Covariance Matrix Sensing}
	%\vspace{-5ex}
	\subsection{Rank Reduction by Permutation} %Structured Low-Rank \tj{Covariance Matrix}}
	%We write $\mb T_{t_i}\in\mathbb{C}^{N_t\times N_t}$ as $\mb T_{t_{kl}}$ and $\mb T_{r_i}\in\mathbb{C}^{N_r\times N_r}$ as $\mb T_{r_{kl}}$, where $i=(k-1)L+l$ with $1\leq l\leq L$ and $1\leq k\leq K$.
	%We 
	%\tj{Let us} write $\mb T^t_{{kl}}$ of (\ref{T_tklULA}) and $\mb T^r_{{kl}}$ of (\ref{T_rklULA}) as 
	%\textcolor{green}{[The notation is confusing; need to define $\mb T_i$ not rewrite $\mb T_{kl}$. Can we avoid introducing $i$ while introducing the model? ]}
	Define
	\begin{equation}
	\label{T_ti}
	\mb T^t_{kl}=\frac{\gamma_k}{\sqrt{L}}\widetilde{\mb T}^t_{{kl}}\in\mathbb{C}^{N_t\times N_t}
	%\mb T^t_{{kl}}=\frac{\sqrt{L}}{\gamma_k}\mb T^t_{i}\in\mathbb{C}^{N_t\times N_t}
	\end{equation}
	and 
	\begin{equation}
	\label{T_ri}
	\mb T^r_{kl}=\frac{\gamma_k}{\sqrt{L}}\widetilde{\mb T}^r_{{kl}}\in\mathbb{C}^{N_r\times N_r}
	%\mb T^r_{{kl}}=\frac{\sqrt{L}}{\gamma_k}\mb T^r_{i}\in\mathbb{C}^{N_r\times N_r},
	\end{equation}
	%as $\frac{\sqrt{L}}{\gamma_k}\mb T_{t_i}\in\mathbb{C}^{N_t\times N_t}$ and $\frac{\sqrt{L}}{\gamma_k}\mb T_{r_i}\in\mathbb{C}^{N_r\times N_r}$, 
	respectively,
	where %$i=(k-1)L+l$ with 
	$1\leq l\leq L$ and $1\leq k\leq K$.
	Then $\mb R$ of (\ref{R}) can be written compactly as
	%Write a compact form of $\mb R$ as
	%\vspace{-1ex}
	\begin{equation}
	\label{st}
	\mb R=\displaystyle  \sum_{k=1}^K  \sum_{l=1 }^L  {\mb T^t_{kl}\otimes \mb T^r_{kl}} \in\mathbb{C}^{N_{t}N_{r} \times N_{t} N_{r}},
	\end{equation}
	where the summation involves $KL$ terms. Note that  
	$\mb T^t_{kl}$ and $\mb T^r_{kl}$ are Toeplitz-Hermitian. 
	%and correspond to $\frac{\gamma_k}{\sqrt{L}}\mb T_{t_{kl}}$ and $\frac{\gamma_k}{\sqrt{L}}\mb T_{r_{kl}}$ of (\ref{R}), respectively. %The rank of $\mb R$ can be large as $KL$ can be a large number, e.g., $KL=60$, meaning that the dimension of the subspace of $\mb R$ to be estimated can be large. %and when $N_t$ and $N_r$ is large, the estimation task is time and resource consuming. 
	%Notice that the covariance matrix $\mb R$ satisfies the Kronecker expansion product model as it is a summation of $N$ Kronecker products, $\mb R$ can be rearranged into a sum of vector outer products. 
	Denote the following $N_r \times N_r$ submatrix of $\mb R$ as     
	\begin{equation}
	\label{denote1}
	\mb R_{mn} \triangleq [\mb R]_{((m-1)N_{r}+1):mN_{r},((n-1)N_{r}+1):nN_{r}},  
	%&\text{with }l=1,\ldots, N_{t}, k=1,\ldots,N_{t}. 
	\end{equation}
	where % \dc{$mn=(m-1)N_t+n$}, 
	$1\leq m\leq N_t$ and $1\leq n\leq N_t$.   
	Define a permutation operator $\mathcal{P} (\cdot)$ that permutes the $N_tN_r \times N_t N_r$ matrix $\mb R$ into a $N_t^2 \times N_r^2$ matrix 
	\vspace{-1ex}
	\[
	\mb R_{p}=\mathcal{P} (\mb R) 
	\] 
	by stacking each submatrix $\mb R_{mn}$ into a row vector as 
	\[[\mathcal{P} (\mb R)]_{m+(n-1)N_t ,:}=\text{vec}^T(\mb R_{mn})\in \mathbb{C}^{1\times N^2_r }.\]
	%\textcolor{red}{[The meaning of the subscript $mn$ above was not clear. Should be $(m-1)N_t+n$? ]}
	We write 
	\[
	\mb t^t_{kl}=\text{vec}(\mb T^t_{kl})\in \mathbb{C}^{N^2_{t} \times 1},
	\] and 
		\vspace{-0.5ex}
	\[
	\mb t^r_{kl}=\text{vec}(\mb T^r_{kl})\in\mathbb{C}^{N^2_{r} \times 1}.\]
	Then based on the Kronecker product expansion property \cite{KP}, \cite{KPExp}, $\mb R_p$ can be written as a sum of vector outer products  
	%\vspace{-1ex}
	\begin{equation}
	\label{mapP}
	\mb R_p=\displaystyle\sum_{k=1 }^K \sum_{l=1}^L \mb t^t_{kl} (\mb t^r_{kl})^T  \in\mathbb{C}^{N_t^2 \times N_r^2} .
	\end{equation}
	Note that if we have $\mb R_p$, we can obtain $\mb R$ as $\mathcal{P}^{-1}(\mb R_p)$.
	%after permutation, 
	%$\mb R_p$ has different subspaces from $\mb R$. 
	% Take an example of $KL=1$, let $\mb M_{t}$ and $\mb M_{r}$ be the singular matrices of the Hermitian matrices $\mb T_{t}$ and $\mb T_{r}$, respectively. Then the singular matrix of the Hermitian matrix $\mb R$ is $\mb M_{t}\otimes \mb M_{r}$, but the left and right singular matrices of $\mb R_p$ are $\mb M^{\ast}_{t}\otimes \mb M_{t}$ and $\mb M^{\ast}_{r}\otimes \mb M_{r}$, respectively. 
	
	From (\ref{mapP}), the column space of $\mb R_p$ is spanned by $\{\mb t^t_{kl}\}$  %, \mb t^t_{2},\ldots, \mb t^t_{N}\}$ and 
	and the row space of $\mb R_{p}$ is spanned by $\{\mb t^r_{kl}\}$. Recall that $\mb t^t_{kl}=\text{vec}(\mb T^t_{kl})$ and by using the relation between $\mb T^t_{kl}$ and the transmitter array response $\mb a_t(\phi^t_{kl})$ shown in (\ref{T_tklULA}) and (\ref{T_ti}), $\mb t^t_{kl}$ can be written as 
	\begin{align}
	\label{t_ti}\nonumber
	\mb t^t_{kl}&=\frac{\gamma_k}{\sqrt{N_tL}}[\mb a^T_{t}(\phi^t_{kl}),\mathrm{e}^{-j\frac{2\pi}{\lambda_c}d\sin(\phi^t_{kl})}\mb a^T_{t}(\phi^t_{kl}),\ldots,\\\nonumber &\quad\quad\quad\quad\quad\quad \mathrm{e}^{-j(N_t-1)\frac{2\pi}{\lambda_c}d\sin(\phi^t_{kl})}\mb a^T_{t}(\phi^t_{kl})]^T\\
	&=\frac{\gamma_k}{\sqrt{L}}\mb a^{\ast}_t(\phi^t_{kl})\otimes \mb a_t(\phi^t_{kl}),  
	\end{align}
	where $\phi^t_{kl}$ is the azimuth AoD. % and $i=(k-1)L+l$ with $1\leq l\leq L$ and $1\leq k\leq K$.}
	We can see that $\mb t^t_{kl}$ consists of the array response vector $\mb a_{t}(\phi^t_{kl})$ and the column space of $\mb R_p$ is determined by the set
	\[\mathbf{\mathcal{C}}_t=\{\mb a^{\ast}_t(\phi^t_{kl})\otimes \mb a_t(\phi^t_{kl}), 1\leq k\leq K, 1\leq l\leq L\}.\]
	%We first discuss the space spanned by the set
	% \[\bf{\mathcal{S}}_t=\{\mb a_t(\phi^t_{kl}), 1\leq k\leq K, 1\leq l\leq L\}.\]
		\vspace{-0.5ex}
	As introduced earlier, small angular spreads are observed in the mmWave propagation environment, which indicates that the AoDs inside a cluster are closely spaced and their corresponding array response vectors are highly correlated. %and the vectors in $\bf{\mathcal{C}}_t$ are also highly correlated. 
	Therefore, for the $k$-th cluster, though the number of rays $L$ inside can be large, the space spanned by $\{\mb a^{\ast}_t(\phi^t_{kl})\otimes \mb a_t(\phi^t_{kl}), 1\leq l\leq L\}$ may be well approximated by a low-rank space. In addition, since the number of clusters $K$ is generally small (e.g., $K=1$ or $2$), both $\bf{\mathcal{C}}_t$ and $\mb R$ can be low-rank. This is similar to the low-rankness of the mmWave channel $\mb H$, which has been validated by the experimental and simulation results in \cite{mmWave Channel}. % (whose column space is spanned by the transmitter array response vectors $\{\mb a_t(\phi^t_{kl}), 1\leq k\leq K, 1\leq l\leq L\}$), 
	
	% suggest that $\mb H$  has a low-rank approximation that captures almost all its energy.
	%For the $k$-th cluster, though the number of rays $L$ inside it can be large, The space spanned  by  $\bf{\mathcal{S}}_t$ may have a low rank. In addition, since the number of clusters $K$ in the mmWave channel is small (e.g., $K=1$ or $2$), the rank of the space spanned by $\bf{\mathcal{S}}_t$ can be small. Since $\bf{\mathcal{S}}_t$ spans the row space of the channel $\mb H$, $\mb H$ may have a low-rank row subspace. For the same reason, the column subspace of $\mb H$ may also has a small rank, and therefore, $\mb H$ may be approximated as a low-rank matrix. The experiment and simulation results in \cite{mmWave Channel} suggest that $\mb H$  has a low-rank approximation that captures almost all its energy. 
	%We can see that when the AoDs are closely spaced, the vectors in $\bf{\mathcal{C}}_t$ are also highly correlated. 
	%  \textcolor{green}{[Can we discuss directly the rank of $\mb R_p$ rather than let the readers figure out the similarity? }\rh{This explanation is still weak. Should we just delete this and go directly to numerical explanation?]}  
	%Following the same analysis for the low-rank property of $\mb H$, $\mb R_p$ can be approximately low-rank. 
	\vspace{-0.5ex}
	The low-rank property of $\mb R_p$ can be shown numerically. 
	Denote by $r_{\rm{ch}}$ the rank of $\mb R_p$ or $\mb R$, and let $\sigma_1>\sigma_2>\ldots>\sigma_{r_{\rm{ch}}}$ be the  singular values of $\mb R_p$ or $\mb R$.
	We may use
	%\vspace{-1ex}
	\begin{equation}
	\label{energy}
	p_e\overset{\Delta}{=}\frac{\sum^{r_{\rm{sub}}}_{j=1}\sigma^2_j}{\sum^{r_{\rm{ch}}}_{i=1}\sigma^2_i}
	\end{equation}
	to measure the energy captured by a rank-$r_{\rm{sub}}$ approximation of $\mb R_p$ or $\mb R$, where $r_{\rm{sub}}$ is the rank of the subspace of $\mb R_p$ or $\mb R$.
	%%%%%%%%%%%%%%
	\begin{figure}
		\centering
		%	\vspace{-3ex}
		\label{hybrid_sys}
		\includegraphics[width=\columnwidth]{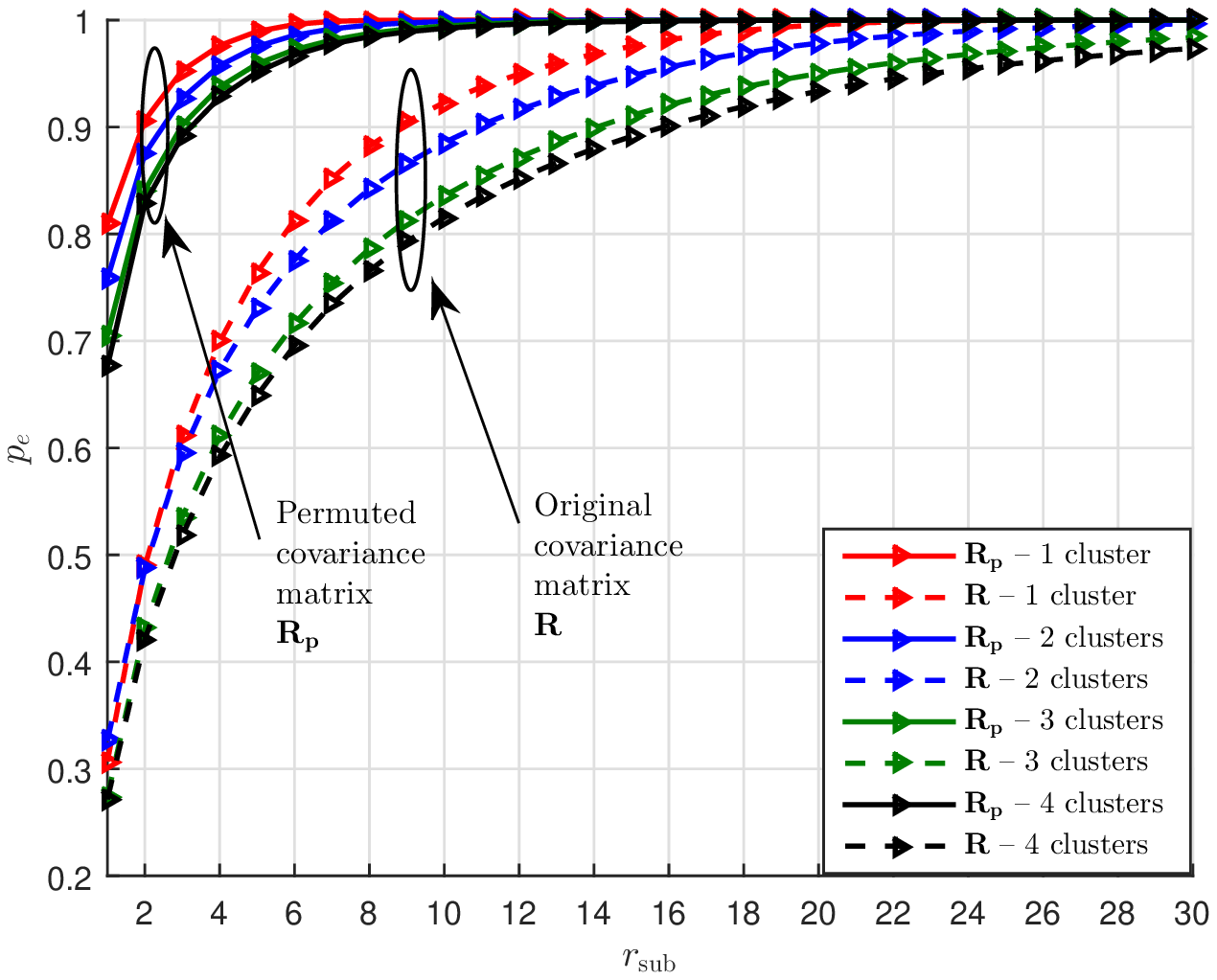}
		\caption{Energy captured by a rank-$r_{\rm{sub}}$ approximation of $\mb R_p$ and $\mb R$}
	\end{figure}
	%%%%%%%%%%%%%%%
	Fig. 2 shows an example of a ULA system with $K=\{1,2,3,4\}, L=30$, $N_t=64$, and $N_r=16$. The covariance matrix $\mb R$ and its permuted version $\mb R_p$ have sizes of $1024\times 1024$ and $4096\times 256$, respectively, which shows that $\mb R_p$ is a taller matrix.
	The horizontal AoDs
	\begin{equation}
	\label{horizontalAoDs}
	\phi^t_{kl}\sim\mathcal{U}(\phi^t_k-\upsilon^t_h,\phi^t_k+\upsilon^t_h), l=1,2,\cdots, L, 
	\end{equation}
	where the center angles $\phi^t_k$ are distributed uniformly in $[0,2\pi]$ and separated by at least one angular spread $\upsilon^t_h=10.2^{\circ}$. Similarly, the horizontal AoAs
	\begin{equation}
	\label{horizontalAoAs}
	\phi^r_{kl}\sim\mathcal{U}(\phi^r_k-\upsilon^r_h,\phi^r_k+\upsilon^r_h), l=1,2,\cdots, L, 
	\end{equation}
	where $\upsilon^r_h=15.5^{\circ}$.
	%The angular spreads at the transmitter side and receiver side are $10.2^{\circ}$ and $15.5^{\circ}$, respectively, and 
	The cluster powers are generated following \cite[Tab. I]{mmWave Channel}. It can be seen from Fig. 2 that for capturing a majority of the total energy, e.g., with $p_e=0.95, 0.99$, the required $r_{\rm{sub}}$ for $\mb R_p$ is generally much smaller than $\text{min}(N^2_t, N^2_r, KL)$ and is also much smaller than that for $\mb R$. 
	%Therefore, $\mb R_p$ can be considered as a low-rank matrix. 
	In the following, we use $r_{p}$ as the $r_{\rm{sub}}$ of $\mb R_p$ and $r_{\rm{R}}$ as the $r_{\rm{sub}}$ of $\mb R$ for a certain $p_e$. 
	%Therefore, we can estimate a rank-$r_p$ matrix $\mb R^r_{p}$ to approximate $\mb R_p$ and then estimate $\mb R$ as $\mathcal{P}^{-1}(\mb R^r_{p})$.
	%We write $\mb R^r_{p}$ as
	%We now rewrite $\mb R_p$ as a rank-$r_{p}$ matrix
	%\begin{equation}
	%\label{R_pre}
	%\mb R^r_{p}=\displaystyle\sum^{r_{p}}_{i=1}\widetilde{\mb t}^t_{i}(\widetilde{\mb t}^r_{i})^T\in\mathbb{C}^{N_t^2 \times N_r^2},
	%\end{equation}
	%where $r_{p}$ is small, $\widetilde{\mb t}^t_{i}\in\mathbb{C}^{N^2_t \times 1}$, and $\widetilde{\mb t}^r_{i}\in\mathbb{C}^{N^2_r \times 1}$. 
	%\textcolor{green}{[Should we discuss why the low-rank approximation still satisfies the structure in (17)?]} 
	As such, $\mb R_p$ may be well approximated as a rank-$r_p$ matrix. One may use low-rank matrix recovery methods, e.g., matrix completion methods, to estimate the best rank-$r_p$ approximation of $\mb R_p$ from a small amount of observations.  
	However, when $N_t$ and $N_r$ are large, which is the case in mmWave communications, the number of parameters required by the rank-$r_p$ approximation of $\mb R_p$, i.e., $(N^2_t + N_t^2)\times r_p$, is still large. Therefore, estimating the subspaces of $\mb R_{p}$ can be computationally expensive.
	\vspace{-1.5ex}
	\subsection{Dimension Reduction By Exploiting the Toeplitz-Hermitian Structure}
	
	Recall that $\mb R$ is block-Toeplitz-Hermitian and $\mb R_p=\mathcal{P}(\mb R)$ is a permutation of $\mb R$. From (\ref{mapP}) and (\ref{t_ti}), we can see that $\mb R_p$ is also specially structured: $\mb R_p$ is the summation of the outer products of $\mb t^t_{kl} $ and $\mb t^r_{kl} $, where $\mb t^t_{kl} $ and $\mb t^r_{kl} $ are the vectorizations of Toeplitz-Hermitian matrices $\mb T^t_{kl} $ and $\mb T^r_{kl} $, respectively. Since the Toeplitz-Hermitian matrix $\mb T^t_{kl} \in\mathbb{C}^{N^2_t\times N^2_t}$ is determined by its first column and first row (its first row is the conjugate transpose of its first column), we can represent $\mb t^t_{kl} $ in terms of the entries in the first column and first row of $\mb T^t_{kl} $. We can represent $\mb t^r_{kl} $ in the same way. 
	Therefore, the total numbers of unknowns in $\mb t^t_{kl} $ and $\mb t^r_{kl} $ are $2N_t-1$ and $2N_r-1$, respectively.  Then we can reduce the problem size of $(N^2_t+N_r^2)\times r_p$ to $2(N_t+N_r-1)\times r_p$. In the following, we show how the problem size can be reduced. 
	First, let us use an example with $N_t=3$ to illustrate the structure of $\mb t^t_{{kl} }$.
	The array response  
	\[\mb a_t(\phi^t_{kl})=\frac{1}{\sqrt{3}}[1, \mathrm{e}^{j \frac{2\pi}{\lambda_c }d\sin(\phi^k_{kl})}, \mathrm{e}^{j2\frac{2\pi}{\lambda_c }d\sin(\phi^k_{kl})}]^T.\] 
	Then according to (\ref{t_ti}), we have
	\begin{equation}\nonumber
	\mb t^t_{kl }=\frac{\sqrt{L}}{\gamma_k}\mb a^{\ast}_t(\phi^t_{kl})\otimes \mb a_t(\phi^t_{kl})=\frac{\sqrt{L}}{3\gamma_k}\begin{bmatrix}
	1\\
	\mathrm{e}^{j\frac{2\pi}{\lambda_c }d\sin(\phi^k_{kl})}\\
	\mathrm{e}^{j2\frac{2\pi}{\lambda_c }d\sin(\phi^k_{kl})}\\
	\mathrm{e}^{-j\frac{2\pi}{\lambda_c }d\sin(\phi^k_{kl})}\\
	1\\
	\mathrm{e}^{j\frac{2\pi}{\lambda_c }d\sin(\phi^k_{kl})}\\
	\mathrm{e}^{-j2\frac{2\pi}{\lambda_c }d\sin(\phi^k_{kl})}\\
	\mathrm{e}^{-j\frac{2\pi}{\lambda_c }d\sin(\phi^k_{kl})}    \\
	1   
	\end{bmatrix}.
	\end{equation}
	%-----------------------------
	We can see that all the $9$ elements in $\mb t^t_{kl}$ can be represented by the elements in $\mb a_t(\phi^t_{kl})$ and $\mb a^{\ast}_t(\phi^t_{kl})$. Now construct a vector \[\mb a_{kl}=\frac{\sqrt{L}}{3\gamma_k}[\mb a^T_t(\phi^t_{kl}), \mathrm{e}^{-j\frac{2\pi}{\lambda_c }d\sin(\phi^k_{kl})}, \mathrm{e}^{-j2\frac{2\pi}{\lambda_c }d\sin(\phi^k_{kl})}]^T\in\mathbb{C}^{5\times 1},\] 
	then $\mb t^t_{kl}$ can be rewritten as
	\begin{equation}\nonumber
	\mb t^t_{kl}=\begin{bmatrix}
	[\mb I_3,\mb 0_{3\times 2}]\\
	[\mb 0_{1\times 3}, 1, 0]\\
	[\mb I_{2},\mb 0_{2\times 3}]\\
	[\mb 0_{1\times 4}, 1]\\
	[\mb 0_{1\times 3}, 1, 0]\\
	[1,\mb 0_{1\times 4}]\\
	\end{bmatrix} \mb a_{kl}.
	\end{equation}
	In fact, $\mb a_{kl}(4)=(\mb a_{kl}(2))^{\ast}$ and $\mb a_{kl}(5)=(\mb a_{kl}(3))^{\ast}$. Therefore, $\mb t^t_{kl}$ can be expressed as a product of a weight matrix and a vector $\mb a_{kl}$. Furthermore, the weight matrix depends only on the structure of the antenna array and is independent of the path angles. 
	
	%In the following we use
	%\begin{equation}
	%\label{mapP_sub}
	%\mb R_p=\displaystyle\sum^{r_{\rm{sub}}}_{i}\mb u_i\mb v^T_i,
	%\end{equation}
	%If $N\ll \text{min}(N^2_{\rm{R}}, N^2_{\rm{T}})$, we can consider that $\mb R_{p}$ has a low rank.%Note that $\text{rank}(\mb R)$ and $\text{rank}(\mb R_p)$ may not be equal. This is because $\text{rank}(\mb R)\leq\displaystyle\sum_i^N \text{rank}(\mb U_i)\text{rank}(\mb V_i)$. 
	%where $\mb u_i$ is a linear combination of $\mb t_{t_i}$ and contains the AoD information, and $\mb v_i$ is a linear combination of $\mb t_{r_i}$ and contains the AoA information. 
	
	%By exploiting the structure of Toeplitz-Hermitian matrices,
	%\textcolor{green}{By assuming that $\widetilde{\mb t}^t_{i}$ and $\widetilde{\mb t}^r_{i}$ have the same structures as those of $\mb t^t_{i}$ and $\mb t^r_{i}$, respectively}, 
	Similarly, for the general cases, we can express  
	$\mb t^t_{kl}$ with a weight matrix $\mb \Gamma_u\in\mathbb{C}^{N^2_t\times (2N_t-1)}$ and a vector {$\mb a_{kl}\in\mathbb{C}^{(2N_t-1)\times 1}$}, and express $\mb t^r_{kl}$ with a weight matrix $\mb \Gamma_v\in\mathbb{C}^{N^2_r\times (2N_r-1)}$ and a vector $\mb b_{kl}\in\mathbb{C}^{(2N_r-1)\times 1 }$. 
	%We require $\mb a_i(1)=(\mb a_i(1))^{\ast}$ and $\mb b_i(1)=(\mb b_i(1))^{\ast}$. 
	We require 
	\[ 
	\mb a_{kl}(x+N_t-1)=(\mb a_{kl}(x))^{\ast}, 2\leq x\leq N_t,\]
	and 
	\[\mb b_{kl}(y+N_r-1)=(\mb b_{kl}(y))^{\ast}, 2\leq y\leq N_r.\]
	%where $2\leq x\leq N_t$ and $2\leq y\leq N_r$.}  
	We then have 
	\begin{equation}
	\label{vecToep}
	\mb t^t_{kl}=\mb \Gamma_{u}\mb a_{kl}, 
	\quad\text{and}\quad \mb t^r_{kl}=\mb \Gamma_{v}\mb b_{kl},
	\end{equation}
	where $\mb \Gamma_u=[\mb \Gamma_{u1},\mb \Gamma_{u2}]$  with $\mb \Gamma_{u1}\in\mathbb{C}^{N^2_{t}\times N_{t}}$, $\mb \Gamma_{u2}\in\mathbb{C}^{N^2_{t}\times (N_{t}-1)}$, and
	%	\vspace{-3ex} 
	\begin{equation}\nonumber
	\mb \Gamma_{u1}=\begin{bmatrix}
	\mb I_{N_{t}}\\
	\mb 0_{1\times N_{t}}\\
	[ \mb I_{ N_{t}-1 },\mb 0_{(N_{t}-1)\times 1}]\\
	\mb 0_{2\times N_{t}}\\
	[ \mb I_{N_{t}-2},\mb 0_{(N_{t}-2)\times 2}]\\
	\mb 0_{3\times N_t}\\
	\vdots\\
	[1,\mb 0_{1\times (N_{t}-1)}]         
	\end{bmatrix},
	\quad \mb \Gamma_{u2}=\begin{bmatrix}
	\mb 0_{N_{t}\times (N_{t}-1)}\\
	\mb e^T_{1}\\
	\mb 0_{N_{t}-1\times (N_{t}-1)}\\
	\mb e^T_{2}\\
	\mb e^T_{1}\\
	\mb 0_{(N_{t}-2)\times (N_{t}-1)}\\
	\mb e^T_{3}\\
	\mb e^T_{2}\\
	\mb e^T_{1}\\
	\vdots\\
	% \widetilde{\mb e}_{N_{t}-1}\\
	%  \widetilde{\mb e}_{N_{t}-2}\\
	%  \vdots\\
	%\widetilde{\mb e}_{1}\\
	\mb 0_{1\times (N_{t}-1)}
	\end{bmatrix}
	\end{equation}
	with $\mb e_i\in\mathbb{C}^{(N_{t}-1)\times 1}$ being a vector whose $i$-th entry is $1$ and other entries are zero. $\mb \Gamma_v$ is constructed similarly as $\mb\Gamma_u$, and $\mb \Gamma_u$ and $\mb \Gamma_v$ are both full-rank. This is because $\mb \Gamma_u$ and $\mb \Gamma_v$ consist of $1$'s and $0$'s, and there is only one $1$ in each row of $\mb \Gamma_u$ and $\mb \Gamma_v$. 
	Therefore, (\ref{mapP}) can be rewritten as 
	\begin{align}
	\label{rewritten}\nonumber
	\mb R_{p}&=\displaystyle \sum_{k=1}^{K} \sum_{l=1}^{L} \mb \Gamma_{u}\mb a_{kl} 
	\mb b^T_{kl} \mb \Gamma^T_{v}\\ \nonumber
	&=\displaystyle\mb \Gamma_{u} \left(  \sum_{k=1}^{K} \sum_{l=1}^{L} \mb a_{kl}
	\mb b^T_{kl} \right)\mb \Gamma^T_{v}\\
	&=\mb \Gamma_{u}\mb C\mb \Gamma^T_{v},
	\end{align}
	where $\mb C= \sum_{k=1}^{K} \sum_{l=1}^{L} \mb a_{kl}
	\mb b^T_{kl} $. % 
	%is a rank-$r_p$ matrix 
	%with
	%\begin{equation}\nonumber
	%\mb A=\begin{bmatrix}
	%\mb a_1&\ldots&\mb a_{N}
	%\end{bmatrix},\quad\text{and}\quad
	%\mb B=\begin{bmatrix}
	%\mb b_1&\ldots&\mb b_{N}\\
	%\end{bmatrix}.
	%\end{equation}
	%Note that $\mb A\in\mathbb{C}^{(2N_{t}-1)\times r_p}$ and $\mb B\in\mathbb{C}^{(2N_{r}-1)\times r_p}$.
	%Then $\mb R_p$ can be written as 
	%\begin{equation}
	%\label{perR}
	%\mb R_p=\mb \Gamma_{u} \mb C\mb \Gamma^T_{v},
	%\end{equation}
	
	As shown above, $\mb R_p$ is approximately low-rank. Since the fixed weight matrices $\mb \Gamma_u$ and $\mb \Gamma_v$ are full-rank, $\mb C$ is approximately low-rank. %Note that $\mb \Gamma_u$ and $\mb \Gamma_v$ are fixed. 
	Hence estimating a low-rank approximation of $\mb R_{p}$ is equivalent to estimating a low-rank approximation of $\mb C$. Note that $\mb C \in \mathbb C^{(2N_t-1) \times (2N_r-1)}$ is much smaller than $\mb R_p \in \mathbb C^{N_t^2 \times N_r^2}$ and this can greatly reduce the complexity of the problem.%Note that $\mb C$ is much smaller than $\mb R_{p}$.
	
	%\textcolor{green}{[It appears to me that the assumptions mentioned above about the structures of the low-rank approximation $\tilde{\mb t}$ are questionable. It seems more nature that low-rank approximation happens to $\mb C$, rather than at the begining; i.e., the low-rankness arises from the property of $\mb C$; that is, the number of terms of summation is still $KL$ but the rank of $\mb C$ is low. ]}
	%	\vspace{-4ex}

	\subsection{Training}
	%\textcolor{green}{[Define or explain first what are ``snapshot'' and ``training steps''?]}
	%	\vspace{-0.5ex}
	We assume that the channel matrix $\mb H$ remains static during a snapshot and suppose we have $T$ snapshots.
	For different snapshots, we assume that the AoAs/AoDs and the fraction power $\gamma_k^2$ remain unchanged, but the small-scale fading gain $g_{kl}\sim\mathcal{CN}(0,\gamma^2_k)$ can change \cite{Joint}. 
	%We assume that the channel remains static during the $S$ training steps per snapshot. 
	Suppose the transmitter sends out $S$ training beams during each snapshot.  
	For the $s$-th training beam of the $t$-th snapshot, we employ the transmitting vector $\mb f_{{t,s}}\in\mathbb{C}^{N_{t}}$ and the receiving matrix $\mb W_{{t,s}}\in\mathbb{C}^{N_{r}\times K_{r}}$.
	Therefore, in each snapshot, after the transmitter sends out $S$ training beams, the receiver receives $SK_r$ symbols and the sampling ratio is  $SK_r/N_rN_t$.	
	We design $\mb f_{t,s}$ and $\mb W_{t,s}$ and their corresponding $\mb F_{\rm{RF}},\mb f_{\rm{BB}},\mb W_{\rm{RF}}$, and $\mb W_{\rm{BB}}$ realizations for the hybrid structure according to the training scheme in \cite[Section III.D]{MCRui}.
	%Following \cite[Section III.D]{MCRui}, $\mb f_{t,s}$  and $\mb W_{t,s}$ are chosen as the column(s) of $\mb X_{\rm{L}}$ and $\mb X_{\rm{R}}$ whose columns are mutually orthogonal, respectively. The corresponding analog/digital design for the hybrid structure can be realized as described in \cite[Section III.D]{MCRui}. 
	For the $s$-th training beam of the $t$-th snapshot, the received signal is%seen at the receiver is
	%\vspace{-1ex}
	\begin{align}
	\label{y_{t,s}} \nonumber
	\mb y_{t,s}&=\mb W^H_{{t,s}}\mb H_t\mb f_{{t,s}}s+\mb W^H_{t,s}\mb n_{t,s}\\
	&=(\mb f^T_{{t,s}}\otimes \mb  W^H_{{t,s}})\text{vec}({\mb H_t})s+\mb W^H_{t,s}\mb n_{t,s}, 
	\end{align}
	where $\mb n_{t,s}$ is the noise vector and ${\mb H_t}$ is the channel matrix at snapshot $t$. Without loss of generality, assume identical training symbols $s=\sqrt{P}$. By setting $\|\mb f_{t,s}\|^2_F=1$, the total transmitting power is $\|\mb f_{t,s} s\|_F^2=P$ and the pilot-to-noise ratio $\rm{(PNR)}$ is defined as 
	\begin{equation} 
	\label{PNR} 
	{\rm{PNR}}=\frac{\|\mb f_{t,s} s\|^2_F}{\sigma^2}, 
	\end{equation}  
	where the noise is assumed to be an additive white Gaussian noise (AWGN) with variance $\sigma^2$. 
	In the $t$-th snapshot and after the transmitter sends out all the $S$ training beams, stack the received signals as%into a column as 
	\begin{align}
	\label{stack}
	\mb y_{t}&=\begin{bmatrix}
	\mb f^T_{{t,1}}\otimes\mb W^H_{t,1}\\
	\mb f^T_{t,2}\otimes\mb W^H_{t,2}\\
	\vdots\\
	\mb f^T_{t,S}\otimes\mb W^H_{t,S}
	\end{bmatrix}\text{vec}(\mb H_t)+\begin{bmatrix}
	\mb W^H_{t,1}\mb n_{t,1}\\
	\mb W^H_{t,2}\mb n_{t,2}\\
	\vdots\\
	\mb W^H_{t,S}\mb n_{t,S}
	\end{bmatrix},\\
	&=\mb P_t\text{vec}(\mb H_t)+\mb n_t \in\mathbb{C}^{SK_r \times 1 },
	\end{align}
	%and $\mb y_t$. 
	where 
	\begin{equation}\nonumber
	\label{Ptnt}
	\mb P_t=\begin{bmatrix}
	\mb f^T_{{t,1}}\otimes\mb W^H_{t,1}\\
	\mb f^T_{t,2}\otimes\mb W^H_{t,2}\\
	\vdots\\
	\mb f^T_{t,S}\otimes\mb W^H_{t,S}
	\end{bmatrix}, \quad\text{and}\quad
	\mb n_t=\begin{bmatrix}
	\mb W^H_{t,1}\mb n_{t,1}\\
	\mb W^H_{t,2}\mb n_{t,2}\\
	\vdots\\
	\mb W^H_{t,S}\mb n_{t,S}
	\end{bmatrix}.
	\end{equation} 
	%\textcolor{green}{[Define $\mb P_t$, $\mb n_t$.]}
	Suppose the trainings are the same for different snapshots, i.e., $\mb f_{1,s}=\mb f_{2,s}=\ldots=\mb f_{T,s}=\mb f_{s}$, $\mb W_{1,s}=\mb W_{2,s}=\ldots=\mb W_{T,s}=\mb W_{s}$. We then have   
	\begin{equation} 
	\mb P=\mb P_1=\ldots=\mb P_T
	=\begin{bmatrix}
	\mb f^T_{1}\otimes\mb W^H_{1}\\
	\mb f^T_{2}\otimes\mb W^H_{2}\\
	\vdots\\
	\mb f^T_{S}\otimes\mb W^H_{S}
	\end{bmatrix}, 
	\end{equation}
	and
	%\rh{The covariance of the received signal is}
	%\vspace{-1ex}
	\begin{equation}
	\label{Covy}
	\mb \Sigma  =\mb P\mb R\mb P^H+\mb \Sigma_n,
	\end{equation}
	where $\mb \Sigma $ and $\mb \Sigma_{n}$ represent the covariance matrices of the received signal $\mb y_t$ and the noise $\mb n_t$, respectively.
	After $T$ snapshots, we can compute the dimension-reduced sample covariance matrix (SCM) of $\mb y_t$ as
	\vspace{-1ex}
	\begin{equation}
	\label{S_stack} 
	\mb S =\frac{1}{T}\displaystyle\sum^{T}_{t=1}\mb y_{t}\mb y^H_{t} \in\mathbb{C}^{SK_r\times SK_r}.
	\end{equation}
	%\textcolor{red}{[This is incorrect as noise and signal cannot be separated in SCM.]}
	%where $\mb h_t=\text{vec}(\mb H_t)$. 
	%and $\mb S\in\mathbb{C}^{SK_r\times SK_r}$.
	%Note that $\mb S$ corresponds to $\mb R$, but we wish to recover $\mb R_p$. Therefore, 
	We permute $\mb S$ into $\mb S_{p} \in\mathbb{C}^{S^2\times K^2_{r}}$ in a similar procedure as $\mb R$ is permuted into $\mb R_p$. 
	%\textcolor{red}{In the following, we will use $\mb R_{{\rm{SCM}}_{p}}=\mathcal{P}_{K_rS}(\mb S)\in\mathbb{C}^{S^2\times K^2_{r}}$. }
	%	\vspace{-1.5ex}
	
	\subsection{Low-Rank Matrix Sensing Problem}
	%	\vspace{-0.5ex}
	We can now formulate the channel covariance estimation problem as a low-rank matrix sensing problem \cite{IMC}:
	\begin{equation}
	\label{MatrixSensing}
	\min_{\widehat{\mb R}_{p}} \text{rank} (\widehat{\mb R}_{p})\quad \text{s.t. } \|\mathcal{A}(\widehat{\mb R}_{p})-\text{vec}(\mb S_p)\|^2_F\leq \zeta^2,
	\end{equation}
	where $\widehat{\mb R}_{p}$ is the estimate of $\mb R_{p}$, $\mathcal{A}: \mathbb{C}^{N^2_t\times N^2_r}\rightarrow \mathbb{C}^{S^2K^2_r \times 1}$ is an appropriate linear map, % \dc{to be detailed soon},  
	and $\zeta^2$ is a constant to account for the fitting error. Replacing $\widehat{\mb R}_{p}$ with (\ref{rewritten}), we can reformulate  (\ref{MatrixSensing}) as
	\begin{equation}
	\label{CMatrix}
	\min_{\widehat{\mb C}} \text{rank} (\widehat{\mb C})\quad \text{s.t. } \|\mathcal{A}(\mb\Gamma_u\widehat{\mb C}\mb\Gamma^T_v)-\text{vec}(\mb S_p)\|^2_F\leq \zeta^2,
	\end{equation}
	where $\widehat{\mb C}$ is the estimate of $\mb C$. 
	
	In general, problem (\ref{CMatrix}) is a nonconvex optimization problem and difficult to solve. In this paper, we solve the relaxed version of problem (\ref{CMatrix}) \cite{GCG-Alt}:
	\begin{equation}
	\label{formulate}
	\min_{\widehat{\mb C}}\phi(\widehat{\mb C})=f(\widehat{\mb C})+\mu\|\widehat{\mb C}\|_{\ast}
	\end{equation}
	where 
	%\vspace{-1ex}
	\begin{equation}
	\label{f_func}
	f(\widehat{\mb C})=\frac{1}{2}\|\mathcal{A}(\mb \Gamma_u\widehat{\mb C}\mb \Gamma^T_v)-\text{vec}(\mb S_p)\|^2_F
	\end{equation}
	and $\mu>0$ is a regularization coefficient.
	%$\|\widehat{\mb C}\|_{\ast}$ is the nuclear norm %(i.e., summation of the singular values) 
	%	of $\widehat{\mb C}$. 
	After some %mathematical 
	manipulations, we have
	$\mathcal{A}(\mb\Gamma_u\widehat{\mb C}\mb\Gamma^T_v)=\mb Q\text{vec}(\widehat{\mb C})$, where 
	\begin{equation}
	\label{Q}
	\mb Q=\begin{bmatrix}
	(\mb f^H_1\otimes\mb f^T_1)\mb \Gamma_u\otimes(\mb W^{\ast}_1\otimes\mb W_1)\mb \Gamma_v\\
	(\mb f^H_1\otimes\mb f^T_2)\mb \Gamma_u\otimes(\mb W^{\ast}_1\otimes\mb W_2)\mb \Gamma_v\\
	\vdots\\
	(\mb f^H_{S}\otimes\mb f^T_{S})\mb\Gamma_u\otimes(\mb W^{\ast}_{S}\otimes\mb W_{S})\mb \Gamma_v\\
	\end{bmatrix},
	\end{equation}
	and $\mb Q\in\mathbb{C}^{S^2K^2_r\times (2N_t-1)(2N_r-1)}$.
	The direct evaluation of $\|\widehat{\mb C}\|_{\ast}$, which is the nuclear norm (i.e., the summation of the singular values) of $\widehat{\mb C}$, is computationally expensive. 
	Following \cite{MCRui}, $\|\widehat{\mb C}\|_{\ast}$ can be written as
	\begin{equation}
	\label{ABnorm}
	\|\widehat{\mb C}\|_{\ast}=\frac{1}{2}\min_{{\mb U},{\mb V}}\{\|{\mb U}\|_F^2+\|{{\mb V}} \|_F^2: \widehat{\mb C}={\mb U}{{\mb V}}^T\}.
	\end{equation}
	Therefore, finding a $\widehat{\mb C}$ to minimize the objective function in (\ref{formulate}) becomes finding a pair of $(\mb U, \mb V)$ to minimize
	\begin{align}
	\label{Main_eq}\nonumber
	\widetilde{\phi}(  \mb U, {\mb V})&\triangleq f( {\mb U}{ {\mb V}}^T)+\frac{1}{2}\mu(\| {\mb U}\|_F^2+\|{ {\mb V}} \|_F^2)\\\nonumber
	&=\frac{1}{2}\|\mb Q\text{vec}(\mb U\mb V^T)-\text{vec}(\mb S_p)\|^2_F\\
	&+\frac{1}{2}\mu(\|\mb U\|^2_F+\|\mb V\|^2_F).
	\end{align}
	%Problem (\ref{Main_eq}) can be solved by the generalized conditional gradient and alternating minimization algorithm (GCG-Alt) in \cite{MCRui}. 
	
	A similar low-rank recovery problem is recently studied in \cite{MCRui} for instantaneous mmWave channel estimation, where a training scheme is designed such that the channel can be estimated by solving a matrix completion (MC) problem.  A generalized conditional gradient and alternating minimization (GCG-Alt) algorithm is developed, which is shown to be able to provide accurate low-rank solutions at a low complexity.  % to solve a matrix completion problem that has the similar formulation as problem (\ref{Main_eq}). 
	%The matrix completion problem solved in \cite{MCRui} has the matrix $\mb Q$ of (\ref{Main_eq}) replaced by a sampling operator $P_{\Omega}$ that samples the entries of the channel.  
	In this work, we adapt the GCG-Alt algorithm to solve (\ref{Main_eq}) for our covariance matrix estimation problem. In the following, we discuss the key steps of the GCG-Alt algorithm for solving (\ref{Main_eq}) and refer the reader to \cite{MCRui} for more detailed treatments.  
	%\textcolor{green}{[Can we say that the GCG-Alt is applied to different problems in [17] and the current paper, i.e., matrix completion and matrix sensing? Is modification of the algorithm needed? Recall reviews.]} 
	%For completeness, we briefly introduce the GCG-Alt algorithm in the following and the details can be found in \cite{MCRui}. 
	%\vspace{-0.5ex}

	The GCG-Alt algorithm consists of a relaxed GCG algorithm and an AltMin algorithm.  Let $\widehat{\mb C}_{k-1}$ be the solution to $\mb C$ at the $(k-1)$-th GCG iteration. %These two algorithms are both iterative. 
	The relaxed GCG algorithm first produces an output
	\begin{equation}
	\label{chatk}
	\widehat{\mb C}_k=(1-\eta_k)\widehat{\mb C}_{k-1}+\theta_k\mb Z_k,
	\end{equation}
	where $\mb Z_k$ is the outer product of the top singular vector pair of $\text{vec}^{-1}(-\nabla f(\widehat{\mb C}_k))$. The calculations of $\text{vec}^{-1}(-\nabla f(\widehat{\mb C}_k))$ and the parameter $\theta_k$ here are different from those in \cite{MCRui}. For problem (\ref{Main_eq}), we calculate $\text{vec}^{-1}(-\nabla f(\widehat{\mb C}_k))$ as 
	\[-\nabla f(\widehat{\mb C}_k)=-(\mb Q^H\mb Q\text{vec}(\widehat{\mb C}_{k-1})-\mb Q\text{vec}(\mb S_p )),\] 
	and the parameter $\theta_k$ as
	\begin{equation}
	\label{theta_k}
	\theta_k=\frac{\mathcal{R}\left(\mb q^H_{z_k}\text{vec}( \mb S_p)-(1-\eta_k)\mb q^H_{z_k}\mb Q\text{vec}(\widehat{\mb C}_k)\right)-\mu}{\mb q^H_{z_k}\mb q_{z_k}},
	\end{equation}
	where $\mb q_{z_k}=\mb Q\text{vec}(\mb Z_k)$ and $\mathcal{R}(\cdot)$ denotes the real part of a number. Since $\widehat{\mb C}_k=\mb U_k\mb V^T_k$, updating $\widehat{\mb C}_k$ is equivalent to updating $\mb U_k=[\sqrt{1-\eta_k}\mb U_{k-1},\sqrt{\theta_k}\mb u_k]$ and $\mb V_k=[\sqrt{1-\eta_k}\mb V_{k-1},\sqrt{\theta_k}\mb v_k]$. Then the obtained $\mb U_k$ and $\mb V_k$ are used as the initial input of the AltMin algorithm, i.e., $\mb U^0_k\leftarrow \mb U_k, \mb V^0_k\leftarrow \mb V_k$. After $I_{a}$ iterations of the AltMin algorithm, update $\mb U_k=\mb U^{I_{a}}_k$ and $\mb V_k=\mb V^{I_{a}}_k$. For completeness, we summarize the GCG-Alt algorithm in Algorithm I.
	After obtaining $\widehat{\mb C}$, we have $\widehat{\mb R}^r_{p}=\mb\Gamma_u\widehat{\mb C}\mb \Gamma^T_u$ and $\widehat{\mb R}=\mathcal{P}^{-1} (\widehat{\mb R}^r_{p})$. %$\widehat{\mb R}=\mathcal{P}^{-1}_{N_rN_t}(\widehat{\mb R}_p)$.
	%%%%
	%	\vspace{-1ex}
	\begin{table}
		\begin{tabular}{ll}\rule{87mm}{.1pt}
			\\
			%	\vspace{-1ex}
			\textbf{Algorithm 1: the GCG-Alt Algorithm for Estimating }\\\textbf{ $\widehat{\mb C}$ of (\ref{formulate}) } \\ 
			\rule[2mm]{87mm}{.1pt} 
		\end{tabular}
		\vspace{-2ex}
		\begin{algorithmic}[1] 
			\State \textbf{Input:} $\text{vec}(\mb S_p ), \mb Q, \mb Q^H\mb Q, \mu,\epsilon, \epsilon_a$ 
			\State \textbf{Initialization:} $ {\bold U}_0=\varnothing , {\bold V}_0=\varnothing,k=0,\epsilon_0=\infty $  
			\While {$\epsilon_{k}>\epsilon$}
			\State $( { {\bold u}}_{k}, { {\bold v}}_{k})\gets$ singular vector pair of $\mb Z_k$%$\text{vec}^{-1}(-\nabla f(\widehat{\bold C}_{k}))$
			\State $k=k+1$
			\State $\eta_k\gets 2/(k+1)$ and determine $\theta_k$ using (\ref{theta_k})
			\State $ {\bold U}_{k}\leftarrow [\sqrt{1-\eta_k} {\bold U}_{k-1},\sqrt{\theta_{k}} {\bold u}_{k}]$
			\State $ {\bold V}_{k}\leftarrow [\sqrt{1-\eta_k} {\bold V}_{k-1},\sqrt{\theta_{k}} \bold v_{k}]$
			\State \textbf{Initialization}:$i=0, \epsilon^{0}_{k}=\infty, ( {\bold U}^0_{k},  {\bold V}^0_{k})\leftarrow ( {\bold U}_{k}, {\bold V}_{k})$
			\While {$\epsilon^i_{k}>\epsilon_a$}
			\State $i=i+1$
			\State update $ {\bold U}^{i}_{k}$ and $ {\bold V}^{i}_{k}$ via the AltMin algorithm \cite{MCRui}
			\State calculate $\epsilon^i_{k}=\frac{\widetilde{\phi}(\mb U^{i-1}_{k},\mb V^{i-1}_k)-\widetilde{\phi}(\mb U^i_k,\mb V^i_k)}{\widetilde{\phi}(\mb U^{i-1}_k,\mb V^i_k)}$ 
			\EndWhile 
			\State $( {\bold U}_{k}, {\bold V}_{k})\leftarrow ( {\bold U}^{i}_{k}, {\bold V}^{i}_{k})$
			\State calculate $\epsilon_{k}=\frac{\widetilde{\phi}(\mb U_{k-1},\mb V_{k-1})-\widetilde{\phi}(\mb U_k,\mb V_k)}{\widetilde{\phi}(\mb U_{k-1},\mb V_{k-1})}$ 
			\EndWhile
			\State \textbf{Output:}  $\widehat{\bold C}=\widehat{\bold C}_{k}= {\bold U}_{k} {\bold V}^T_{k}$
		\end{algorithmic}
		\vspace{-1ex}
		\begin{tabular}{ll}
			\rule[1.5mm]{87mm}{.1pt} 
		\end{tabular}
		%	\vspace{-5ex}
	\end{table}
	%%%%
	%	\vspace{-1ex}
	
	\subsection{Computational Complexity}
	Define a flop as an operation of real-valued numbers. Let $M=SK_r$ be the number of received symbols during each snapshot. 
	%Consider the very recently proposed CS-based  channel covariance estimator DCOMP \cite{CS} whose number of grid points for the AoD and AoA are $G_t$ and $G_r$, respectively. Let $L_p$ be the number of paths for the DCOMP estimator as it assumes $L_p$ is known as a prior. The computational complexity of the DCOMP estimator is about $8TL_pG_tG_r(M^2+M)$. 
	%
	Following the computational complexity analysis in \cite{MCRui}, the computational complexity of the GCG-Alt estimator is about $8r_{\rm{est}}(I_ar_{\rm{est}}+I_a+1)(2N_t-1)^2(2N_r-1)^2+8/3I_ar_{\rm{est}}(r_{\rm{est}}+1)(2r_{\rm{est}}+1)(2N_t-1)(2N_r-1)(N_r+N_t-1)+I_ar^2_{\rm{est}}(r_{\rm{est}}+1)^2((2N_r-1)^3+(2N_t-1)^3)+16r_{\rm{est}}(2N_r-1)(2N_t-1)M^2$, where $I_a$ is the number of iterations of the AltMin algorithm and $r_{\rm{est}}$ is the estimated rank of $\widehat{\mb C}$ by the GCG-Alt estimator. 
	Later in Section IV, we show the computational complexity of the GCG-Alt estimator with specific examples.  
	%
	%Based on our observation, $I_a\leq2$ and $r_{\rm{est}}$ is around $3$ or $4$. If $G_t=128, G_r=32, N_t=64, N_r=16, S=32, K_r=4, T=20, L_p=18, r_{\rm{est}}=4$, and $ I_a=2$,  then the DCOMP estimator requires approximately $2\times 10^{11}$ flops while the GCG-Alt estimator requires approximately $1.15\times 10^{10}$ flops. 
	%
	%The total computational complexity for the DCOMP is approximately 17.4 times that of the GCG-Alt estimator. 
	
	\subsection{Extension to the USPA System}
	%We now discuss how to implement our proposed method in USPA systems.
	We now follow the same process introduced in Section III. A-D to estimate the channel covariance matrix for USPA systems. %estimation problem as a low-rank matrix sensing problem. However, s
	To account for the different array structure of the USPA, the weight matrices of (\ref{vecToep}) are redesigned.   
	For a $\sqrt{N}_a\times \sqrt{N}_a$ USPA placed on the $yz$ plane with distance $d$ between adjacent antennas, the array response is 
	\begin{equation}
	\label{kronA}
	\mb a(\phi_{kl}, \theta_{kl})=\mb a_y(\phi_{kl}, \theta_{kl})\otimes \mb a_z(\theta_{kl}),
	\end{equation}
	where
	\[\mb a_y(\phi_{kl},\theta_{kl})=\frac{1}{N_a^{\frac{1}{4}}}[1, \mathrm{e}^{j \frac{2\pi}{\lambda_c}d\sin(\phi_{kl})\sin(\theta_{kl})},\]
	\[\cdots,\mathrm{e}^{j(\sqrt{N_a}-1) \frac{2\pi}{\lambda_c}d\sin(\phi_{kl})\sin(\theta_{kl})} ]^T\]
	is the array response along the $y$ axis  and 
	\[\mb a_z(\theta_{kl})=\frac{1}{N_a^{\frac{1}{4}}}[1, \mathrm{e}^{j \frac{2\pi}{\lambda_c}d\cos(\theta_{kl})},
	\cdots,\mathrm{e}^{j(\sqrt{N_a}-1)\frac{2\pi}{\lambda_c}d\cos(\theta_{kl})} ]^T  \] is the array response along the $z$ axis. 
	%Then $\mb t_{t_i}$ of (\ref{t_ti}) becomes 
	%\begin{align}
	%\label{USPAt_ti}\nonumber
	%\mb t_{t_i}&=\frac{\sqrt{L}}{\gamma_k}\mb a^{\ast}_t(\phi^t_{kl},\theta^t_{kl})\otimes \mb a_t(\phi^t_{kl},\theta^t_{kl})\\ \nonumber
	%&=\frac{\sqrt{L}}{\gamma_k}(\mb a_{t_y}(\phi^t_{kl},\theta^t_{kl})\otimes \mb a_{t_z}(\theta^t_{kl}))^{\ast}\\
	%&\quad\quad\otimes (\mb a_{t_y}(\phi^t_{kl},\theta^t_{kl})\otimes \mb a_{t_z}(\theta^t_{kl})). 
	%\end{align}
	%Comparing (\ref{t_ti}) and (\ref{USPAt_ti}), we can see that $\mb a_{t}(\phi^t_{kl})$ of (\ref{t_ti}) corresponds to  $(\mb a_{t_y}(\phi^t_{kl},\theta^t_{kl})\otimes \mb a_{t_z}(\theta^t_{kl}))$ of (\ref{USPAt_ti}). Since $\mb t_{t_i}$ of (\ref{t_ti}) can be represented as a product of a weight matrix and a vector consists of $\mb a_t(\phi^t_{kl})$, $\mb t_{t_i}$ of (\ref{USPAt_ti}) can be constructed as a weight matrix and a vector consists of $(\mb a_{t_y}(\phi^t_{kl},\theta^t_{kl})\otimes \mb a_{t_z}(\theta^t_{kl}))$.
	%-------------------------------------
	We design the weight matrices by examining the structure of $\widetilde{\mb T}^t_{kl}$ defined in (\ref{T_tkl}) which is written as  
	\begin{align}
	\label{USPAT_ti}\nonumber
	\widetilde{\mb T}^t_{{kl}}&=\mb a^{\ast}_t(\phi^t_{kl},\theta^t_{kl})\mb a^T_t(\phi^t_{kl},\theta^t_{kl})\\\nonumber
	&= \left(\mb a_{t_y}(\phi^t_{kl},\theta^t_{kl})\otimes \mb a_{t_z}(\theta^t_{kl})\right)^{\ast} \left(\mb a_{t_y}(\phi^t_{kl},\theta^t_{kl})\otimes \mb a_{t_z}(\theta^t_{kl})\right)^T
	%\\ &=\left(\mb a^{\ast}_{t_y}(\phi^t_{kl},\theta^t_{kl})\mb a^T_{t_y}(\phi^t_{kl},\theta^t_{kl})\right)\otimes \left(\mb a^{\ast}_{t_z}(\theta^t_{kl})\mb a^T_{t_z}(\theta^t_{kl})\right),
	\end{align}
	where $\mb a_{t_{y}}(\phi^t_{kl},\theta^t_{kl})$ and $\mb a_{t_{z}}(\theta^t_{kl})$ are the transmitter array response vectors along the $y$ axis and $z$ axis, respectively. 
	Note that $\widetilde{\mb T}^t_{{kl}}$ is block-Toeplitz-Hermitian. 
	Let 
	\[ \widetilde{\mb T} ^y_{{kl}}= \mb a^{\ast}_{t_y}(\phi^t_{kl},\theta^t_{kl})\mb a^T_{t_y}(\phi^t_{kl},\theta^t_{kl}) \in\mathbb{C}^{\sqrt{N_t}\times \sqrt{N_t}}\] 
	and 
	\[\widetilde{\mb T}^z_{{kl}}= \mb a^{\ast}_{t_z}(\theta^t_{kl})\mb a^T_{t_z}(\theta^t_{kl}) \in\mathbb{C}^{\sqrt{N_t}\times \sqrt{N_t}},\] 
	we can verify that %have 
	\begin{equation}
	\label{T_tklUSPACom}
	\widetilde{\mb T}^t_{{kl}}=\widetilde{\mb T}^y_{{kl}}\otimes \widetilde{\mb T}^z_{{kl}},
	\end{equation}
	and $\widetilde{\mb T}^y_{{kl}}$ and $\widetilde{\mb T}^z_{{kl}}$ are Toeplitz-Hermitian. Then for the USPA, $\mb T^t_{kl}$ of (\ref{T_ti}) can be written as  
	\begin{equation}
	\label{T_tiUSPA}
	\mb T^t_{kl}=\frac{\gamma_k}{\sqrt{L}} \widetilde{\mb T}^y_{{kl}}\otimes \widetilde{\mb T}^z_{{kl}}. 
	\end{equation}
	%where $i=(k-1)L+l$ with $1\leq l\leq L$ and $1\leq k\leq K$. 
	In Section III. B, we have expressed $\mb t^t_{kl}=\text{vec}^{-1}(\mb T^t_{kl})$, where $\mb T^t_{kl}$ of (\ref{T_ti}) is Toeplitz-Hermitian matrix, in terms of a weight matrix and a vector. We have similar expressions %can also construct 
	for the vectorizations of the Toeplitz-Hermitian matrices $\widetilde{\mb T}^y_{{kl}}$ and $\widetilde{\mb T}^z_{{kl}}$. % in terms weight matrices and vectors. Suppose we have 
	Let
	\begin{align}
	\label{vecT_ykl}\nonumber
	\widetilde{\mb t}^y_{kl}&=\text{vec}(\widetilde{\mb T}^y_{{kl}})\\
	&=\mb \Gamma_y \mb a^y_{kl}
	\end{align}
	where $\mb \Gamma_y\in\mathbb{C}^{N_t\times (2\sqrt{N_t}-1)}$ is the weight matrix and $\mb a^y_{kl}\in\mathbb{C}^{(2\sqrt{N_t}-1) \times 1}$, and
	\begin{align}
	\label{vecT_zkl}\nonumber
	\widetilde{\mb t}^z_{kl}&= \text{vec}(\widetilde{\mb T}^z_{{kl}})\\
	&=\mb \Gamma_{z}\mb a^z_{kl}
	\end{align}
	where $\mb \Gamma_{z}\in\mathbb{C}^{N_t \times 
		(2\sqrt{N_t}-1)}$ is the weight matrix and $\mb a^z_{kl}\in\mathbb{C}^{(2\sqrt{N_t}-1) \times 1}$. 
	Let 
	\[  \mb\Gamma^{(a)}_{y}=[\mb \Gamma_{y}]_{1+(a-1)\sqrt{N_t}:a\sqrt{N_t},:, } \; 1\leq a\leq \sqrt{N_t},
	\]  %with $1\leq a\leq \sqrt{N_t}$, 
	and 
	\[ 
	\mb \Gamma^{(b)}_{z}=[\mb \Gamma_z]_{1+(b-1)\sqrt{N_t}:b\sqrt{N_t},:}, \; 1\leq b\leq \sqrt{N_t}.
	\] %with $1\leq b\leq \sqrt{N_t}$. 
	By exploring the matrix vectorization process, we have 
	\begin{align}
	\label{t_tiUSPA}\nonumber
	\mb t^t_{kl}&= \text{vec}(\mb T^t_{kl})\\
	%&=\frac{\gamma_k}{\sqrt{L}}\mb \Gamma_u(\mb a^y_{i}\otimes \mb a^z_{i})\\
	&=\mb \Gamma_u \mb a_{kl},
	\end{align}
	where 
	\begin{equation} 
	\label{GammaUSPA}
	\mb \Gamma_u
	=\begin{bmatrix}
	\mb \Gamma^{(1)}_{y}\otimes \mb \Gamma^{(1)}_{z}\\
	\mb \Gamma^{(1)}_{y}\otimes \mb \Gamma^{(2)}_{z}\\
	\vdots\\
	\mb \Gamma^{(1)}_{y}\otimes \mb \Gamma^{(\sqrt{N_t})}_{z}\\
	\mb \Gamma^{(2)}_{y}\otimes \mb \Gamma^{(1)}_{z}\\
	\mb \Gamma^{(2)}_{y}\otimes \mb \Gamma^{(2)}_{z}\\
	\vdots\\
	\mb \Gamma^{(\sqrt{N_t})}_{y}\otimes \mb \Gamma^{(\sqrt{N_t})}_{z}
	\end{bmatrix}\in\mathbb{C}^{N^2_t\times (2\sqrt{N_t}-1)^2} 
	\end{equation}
	is the weight matrix and 
	\[\mb a_{kl}=\frac{\gamma_k}{\sqrt{L}}(\mb a^y_{kl}\otimes\mb a^z_{kl})\in\mathbb{C}^{(2\sqrt{N_t}-1)^2 \times 1}\] 
	is a vector.  
	Then for the USPA system, $\mb \Gamma_u$ of (\ref{vecToep}) becomes (\ref{GammaUSPA}) and $\mb \Gamma_v$ of (\ref{vecToep}) is constructed similarly as (\ref{GammaUSPA}); the sizes of vectors $\mb a_{kl}$ and $\mb b_{kl}$ of (\ref{vecToep}) have changed: $\mb a_{kl}\in\mathbb{C}^{(2\sqrt{N_t}-1)^2 \times 1}$ and $\mb b_{kl}\in\mathbb{C}^{(2\sqrt{N_r}-1)^2 \times 1}$, and consequently, the size for matrix $\mb C$ of (\ref{rewritten}) has changed: {$\mb C\in\mathbb{C}^{(2\sqrt{N_t}-1)^2\times (2\sqrt{N_r}-1)^2}$}. After obtaining the weight matrices, we can follow the processes in Section III. C-D to estimate $\mb C$ and then have the channel covariance matrix estimated as $\widehat{\mb R}=\mathcal{P}^{-1}(\mb\Gamma_u\widehat{\mb C}\mb\Gamma^T_v)$.
	%%%%%%%
	%	\vspace{-2ex}

	\section{Simulations}
	We now evaluate the performance of our proposed design for fully connected hybrid
	transceivers with the ULA and USPA.
	\subsection{The ULA system}
	We assume a carrier frequency of $f_c=28$ GHz.  For the ULA system,  $N_t=64, N_r=16, K_t=16$, and $K_r=4$. The number of clusters $K=\{1, 2\}$ and there are $L=30$ rays in each cluster. 
	%The transmitting angular spread is $10.2^{\circ}$, the receiving angular spread is $15.5^{\circ}$, 
	The horizontal AoDs and AoAs are generated as (\ref{horizontalAoDs}) with $\upsilon^t_h=10.2^{\circ}$ and as (\ref{horizontalAoAs}) with $\upsilon^r_h=15.5^{\circ}$, respectively. 
	The cluster powers are generated following \cite[Tab. I]{mmWave Channel}. We compare the GCG-Alt estimator with the DCOMP estimator in \cite{CS}, which has varying receiving processing matrix $\mb W_{t,s}$ and transmitting processing vector $\mb f_{t,s}$ during training and has the best performance among other estimators in \cite{CS}. The DCOMP estimator needs a dictionary matrix with $G_t$ grid points that is associated with AoD and a dictionary matrix with $G_r$ grid points that is associated with AoA. Let $L_p$ be the number of paths in the channel, the DCOMP estimator assumes that $L_p$ is known. 
	For the DCOMP estimator, we set $G_t=2N_t=128, G_r=2N_r=32$, and $L_p=r_{\rm{R}}$.
	Based on Fig. 2, for $p_e = 0.99$, $r_{\rm{R}} = 18$ and $24$ for $K = 1$ and $2$, respectively.
	For the GCG-Alt estimator, we set $\mu=\sigma^2$, $\epsilon=0.003$, and $\epsilon_a=0.1$. 
	The performance metric $\eta$ \cite{CS} \[\eta=\frac{\text{tr}(\widehat{\mb M}^H\mb R\widehat{\mb M})}{\text{tr}(\mb M^H\mb R\mb M)} \] 
	is used to measure how close the subspace of $\widehat{\mb R}$ is to the subspace of $\mb R$, where $\widehat{\mb M}\in\mathbb{C}^{N_tN_r\times r_{\rm{R}}}$ and $\mb M\in\mathbb{C}^{N_tN_r\times r_{\rm{R}}}$ are the singular vector matrices of $\widehat{\mb R}$ and $\mb R$, respectively. 
	We also use the average of the normalized mean square error
	\[
	{\rm{NMSE}}=\frac{\|\widehat{\mb R}-\mb R\|^2_F}{\|\mb R\|^2_F}
	\]
	to measure their performance. 
	
	We set ${\rm{PNR}}=10$ dB and the number of training beams $S=32$, and compare the GCG-Alt estimator with the DCOMP estimator under different $T$. With $S=32$ per snapshot, the sampling ratio at each snapshot is $SK_r/N_rN_t=12.5\%$. 
	The comparison result shown in Fig. 3
	%We also use ${\rm NMSE} =\|\widehat{\mb R}-\mb R\|^2_F/\|\mb R\|^2_F$ to measure their performance. 
	% The $\eta$ evaluation is shown in Fig. 2 and the NMSE evaluation is shown in Fig. 3. 
	%The performance is shown in Fig. 2 and 3.  
	suggests that when the sampling ratio per snapshot is $12.5\%$, our proposed estimator requires fewer snapshots to obtain a $\widehat{\mb R}$ whose subspace is close to that of $\mb R$, as compared to the DCOMP estimator. 
	The $\rm{NMSE}$ result shown in Fig. 4 suggests that our proposed GCG-Alt estimator can obtain a more accurate covariance matrix estimate. 
	
	We also compare the computational complexity of the GCG-Alt estimator and the DCOMP estimator. The computational complexity of the DCOMP estimator is about $8TL_pG_tG_r(M^2+M)$ flops, where $M=SK_r$. For the GCG-Alt estimator, based on our observations, the number of iterations of the AltMin algorithm $I_a\leq2$, the estimated rank $r_{\rm{est}}\approx 4$ when $K=1$ and $r_{\rm{est}}\approx 5$ when $K=2$. Fig. 5 shows the comparison results with different $T$. We can see that the computational complexity of the GCG-Alt estimator is lower than the DCOMP estimator. Also, the computational complexity of the GCG-Alt estimator does not increase as $T$ increases. This is because we use $\mb S_p\in\mathbb{C}^{S^2\times K^2_r}$, which is the permutation of the SCM of $\mb y_t$ shown in (\ref{S_stack}), and its size is irrelevant to $T$.
	
	\begin{figure}
		\centering
		\label{Eta}
		\includegraphics[width=\columnwidth]{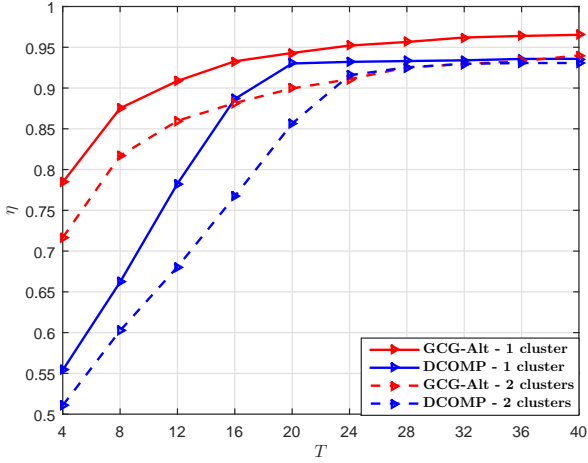}
		\caption{Comparison of $\eta$ of the GCG-Alt estimator and the DCOMP estimator under the ULA system, where $N_t=64, N_r=16, \rm{PNR}=10$ dB, and $S=32$.}
		%\vspace{-2ex}
	\end{figure}
	\begin{figure}
		\centering
		\label{NMSE}
		\includegraphics[width=\columnwidth]{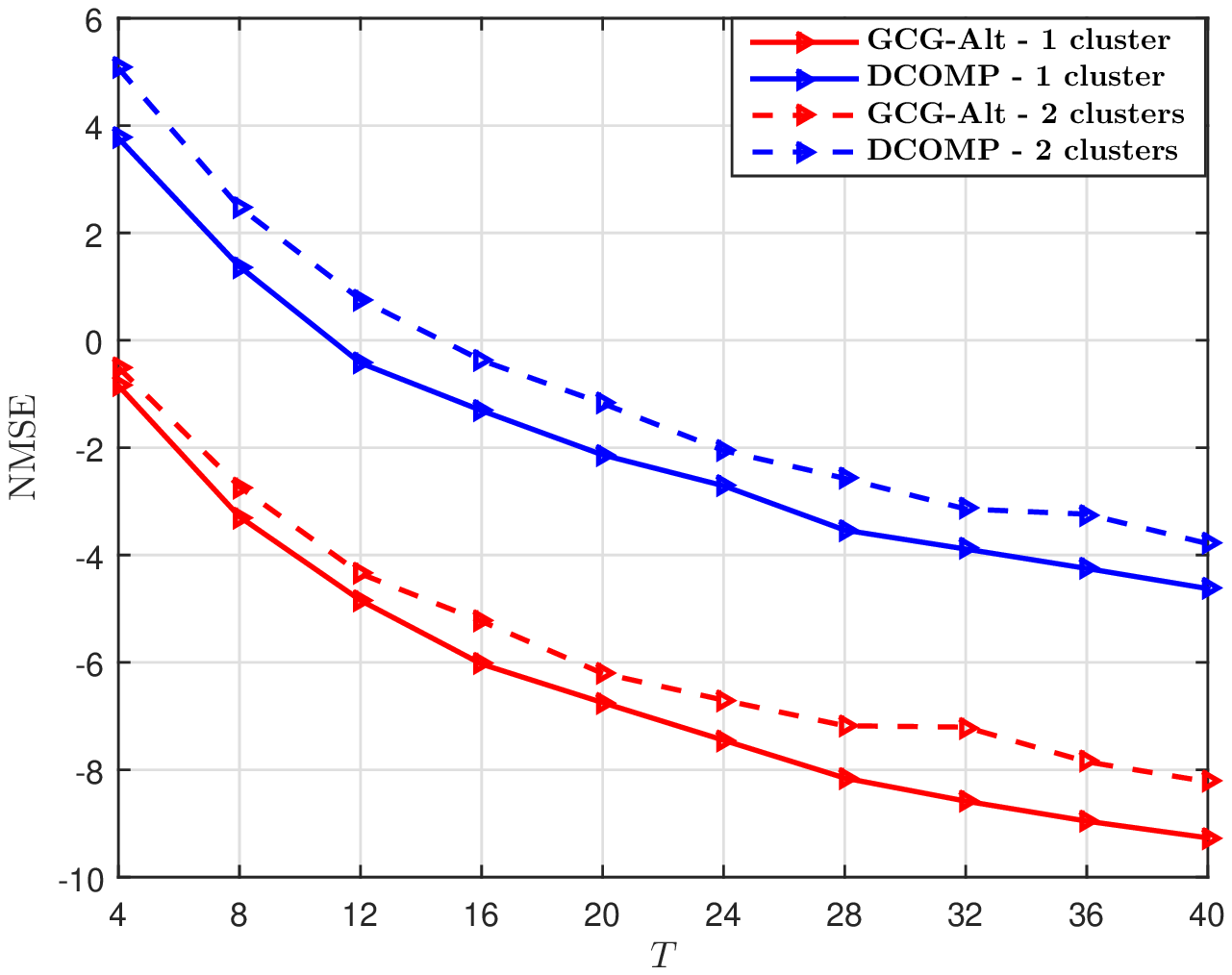}
		\caption{Comparison of $\rm{NMSE}$ of the GCG-Alt estimator and the DCOMP estimator under the ULA system, where $N_t=64, N_r=16, \rm{PNR}=10$ dB, and $S=32$.}
		%\vspace{-2ex}
	\end{figure}
	\begin{figure}
		\centering
		\label{Flops_ULA}
		\includegraphics[width=\columnwidth]{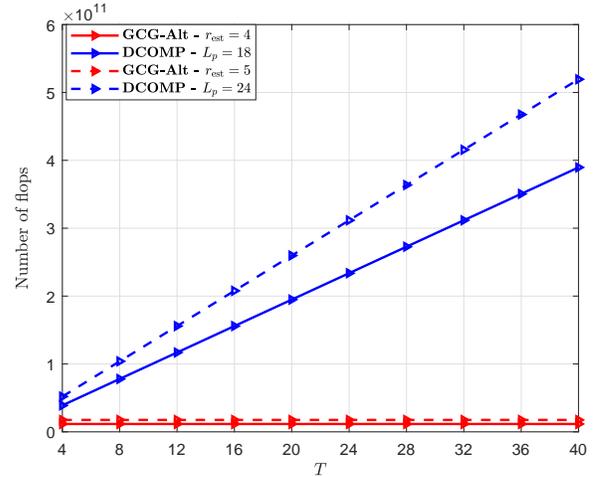}
		\caption{Complexity comparison of the GCG-Alt estimator and the DCOMP estimator under the ULA system, where $N_t=64, N_r=16, \rm{PNR}=10$ dB, and $S=32$.}
	\end{figure}
	\begin{figure}
		%	\vspace{-1ex}
		\centering
		\label{Eta_M_T}
		\includegraphics[width=\columnwidth]{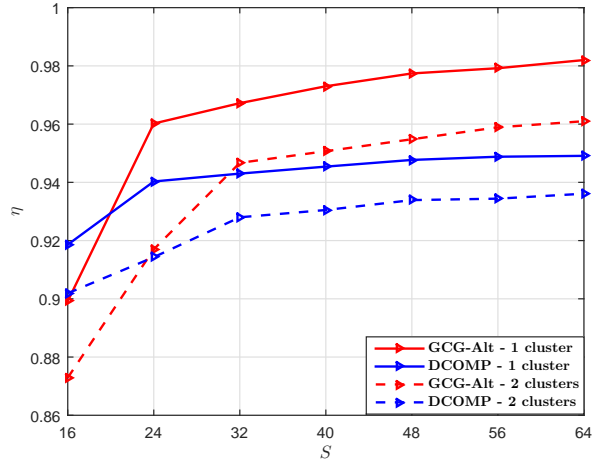}
		\caption{Comparison of $\eta$ of the GCG-Alt estimator and the DCOMP estimator under the ULA system, where $N_t=64, N_r=16, \rm{PNR}=10$ dB, and $T=40$.}
		%	\vspace{-3ex}
	\end{figure}
	
	Then we set %${\rm{PNR}}=10$ dB and 
	the number of snapshots $T=40$, and compare the GCG-Alt estimator with the DCOMP estimator under different $S$. The result shown in Fig. 6 suggests that when $T=40$, the GCG-Alt estimator can obtain a more accurate subspace estimation than the DCOMP estimator when the number of training beams $S\geq 24$ per snapshot. Note that $S=24$ corresponds to a sampling ratio of $9.375\%$ per snapshot.
	
	The GCG-Alt estimator explores both the Kronecker structure and the block-Toeplitz-Hermitian structure of $\mb R$ while the DCOMP estimator only considers the Hermitian structure of $\mb R$, so the GCG-Alt estimator can reach an accurate subspace estimation of $\mb R$ with fewer snapshots. We use the same training for different snapshots while the DCOMP estimator uses different trainings per snapshot (i.e., varying $\mb W_{t,s}$ and $\mb f_{t,s}$ ). % therefore, w
	When $S$ is small (e.g., $S\leq16$), the DCOMP estimator outperforms the GCG-Alt estimator. However, the GCG-Alt estimator performs better when $S$ becomes larger (e.g., $S\geq 24$). Note that for the DCOMP estimator, estimating more paths (i.e., $L_p$ is large) yields better performance, but its computational complexity also increases. 
	%and Fig. 3 shows that our proposed estimator can obtain a more accurate covariance matrix estimate. % than the DCOMP estimator.  
	%

	\begin{figure}
		\centering
		\label{EtaUSPA}
		\includegraphics[width=\columnwidth]{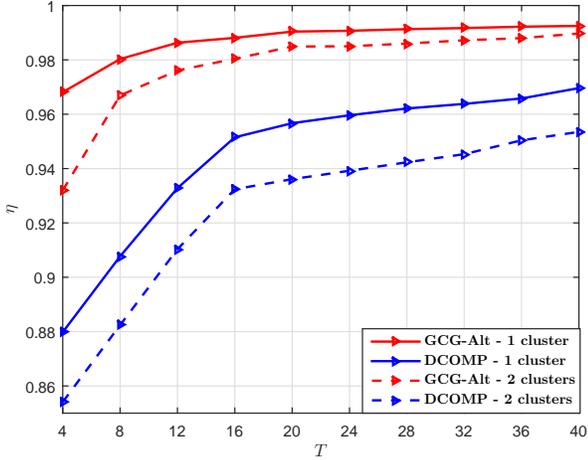}
		\caption{Comparison of $\eta$ of the GCG-Alt estimator and the DCOMP estimator under the USPA system, where $N_t=64, N_r=16, \rm{PNR}=10$ dB, and $S=32$.}
	\end{figure}
	
	\begin{figure}
		\centering
		\label{EtaUSPA_S}
		\includegraphics[width=\columnwidth]{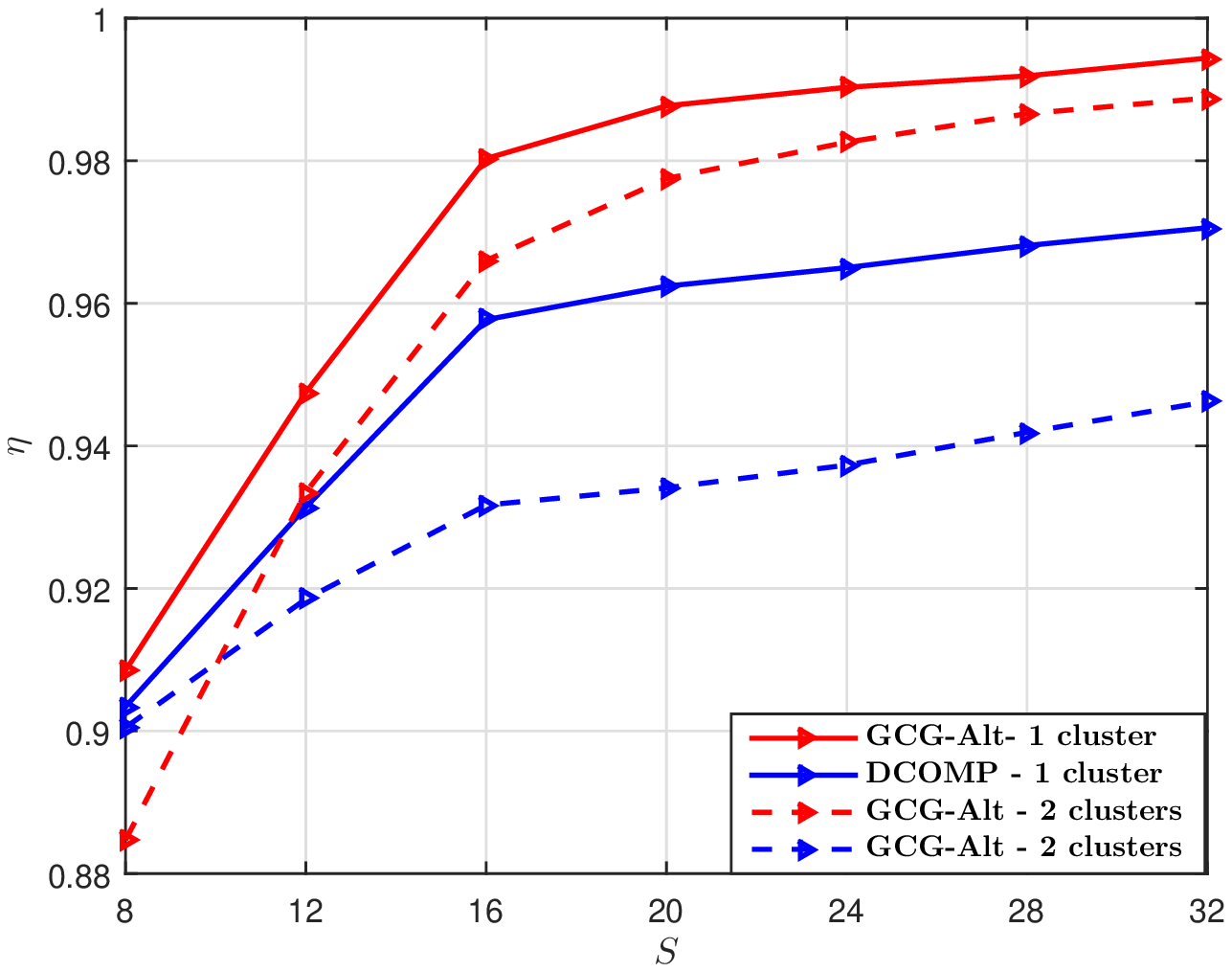}
		\caption{Comparison of $\eta$ of the GCG-Alt estimator and the DCOMP estimator under the USPA system, where $N_t=64, N_r=16, \rm{PNR}=10$ dB, and $T=40$.}
	\end{figure}
	
	\begin{figure}
		\centering
		\label{EtaUSPA_S}
		\includegraphics[width=\columnwidth]{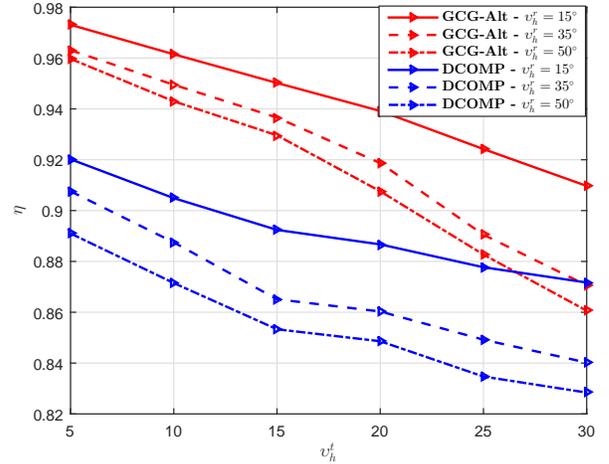}
		\caption{Comparison of $\eta$ of the GCG-Alt estimator and the DCOMP estimator under the USPA system, where $N_t=64, N_r=16, K=1, \rm{PNR}=10$ dB, $S=16, T=16$, $\upsilon^t_v=0^{\circ}$, and $\upsilon^r_v=6^{\circ}$.}
	\end{figure}
	
	\subsection{The USPA system}
	We next consider the system with the USPA at the transmitter and receiver. The parameters $f_c, K, L, \phi^t_{kl}, \phi^r_{kl}$ are assumed the same as in the ULA system. The transmitter has an $8\times 8$ USPA (i.e., $N_t=64$) and $K_t=16$ RF chains, and the receiver has a $4\times 4$ USPA (i.e., $N_r=16$) and $K_r=4$ RF chains. 
	We assume the elevation AoD angular spread $\upsilon^t_v=0^{\circ}$ and the elevation AoA angular spread $\upsilon^r_v=6^{\circ}$ based on the measurement results in \cite{mmWave Channel}. The elevation AoDs and AoAs are distributed as
	\[
	\theta^t_{kl}\sim\mathcal{U}(\theta^t_k-\upsilon^t_v,\theta^t_k+\upsilon^t_{v}),
	\] 
	\[
	\theta^r_{kl}\sim\mathcal{U}(\theta^r_k-\upsilon^r_v,\theta^r_k+\upsilon^r_{v}),
	\]
	with the elevation center angles $\theta^t_k$ and $\theta^r_k$ being generated in the same manner as the azimuth center angles in the ULA system. 
	For the DCOMP estimator, we set $G_t=2\sqrt{N_t}\times 2\sqrt{N_t}=256$ and $G_r=2\sqrt{N_r}\times 2\sqrt{N_r}=64$. The parameters $L_p, \mu ,\epsilon,\epsilon_a$ for the GCG-Alt estimator and DCOMP estimator are the same as in the ULA system.
	
	We set ${\rm{PNR}}=10$ dB. The performance comparison with $S=32$ under different $T$ is shown in Fig. 7 and the performance comparison with $T=40$ under different $S$ is shown in Fig. 8.  We can see that both of the GCG-Alt estimator and the DCOMP estimator achieve higher $\eta$ for the USPA system. One reason for this is that the USPA system has lower resolution than the ULA system in the azimuth direction even though they have the same number of transmitter and receiver antennas. For the USPA system, the azimuth AoD is resolved  by an $\sqrt{N_t}=8$-element antenna array and the azimuth AoA is resolved by a $\sqrt{N_r}=4$-element antenna array; while for the ULA system, the azimuth AoD is resolved  by a $N_t=64$-element antenna array and the azimuth AoA is resolved  by a $N_r=16$-element antenna array. Therefore, for the same angular spread, the USPA system resolves fewer paths than the ULA system, which results in a lower rank. 
	
	We also show the effects of angular spreads on the performance of the estimators. 
	We set $\upsilon^t_v=0^{\circ}, \upsilon^r_v=6^{\circ}, K=1, {\rm{PNR}}=10$ dB, $S=16$, and $T=16$. The estimators' performance under different angular spreads for the azimuth AoD/AoA (i.e., different $\upsilon^t_h$ and $\upsilon^r_h$) shown in Fig. 9 suggests that the estimators achieve lower $\eta$ when $\upsilon^t_h$ and $\upsilon^r_h$ are larger.
	
	\section{Conclusions}
	We have formulated the channel covariance estimation problem for hybrid mmWave systems as a structured low-rank matrix sensing problem by exploiting Kronecker product expansion and the structures of the ULA/USPA. The formulated problem has a reduced dimensionality and is solved by using a low-complexity  GCG-Alt algorithm. The computational complexity analysis and numerical results suggest that our proposed method is effective in estimating the mmWave channel covariance matrix.
	
	%We have formulated the channel covariance estimation problem as a structured low-rank matrix sensing problem via Kronecker product expansion and propose to use the GCG-Alt algorithm to solve it. The computational complexity analysis and numerical result suggest that our proposed method has lower computational complexity and better performance than the DCOMP estimator.


\begin{thebibliography}{00}
		
		
		\bibitem{SurveyI} Y. Niu, Y. Li, D. Jin, L. Su, and A. V. Vasilakos,
		``A survey of millimeter wave communications (mmWave) for 5G: opportunities and challenges,"
		\emph{Wireless Networks}, vol. 21, no. 8, pp. 2657--2676, Apr. 2015.
		
		\bibitem{SurveyII} X. Wang, L. Kong, F. Kong, F. Qiu, M. Xia, S. Arnon, and G. Chen
		``Millimeter wave communication: A comprehensive survey,"
		\emph{IEEE Commun. Surveys Tuts. 
		}, vol. 20, no. 3, pp. 1616--1653, Aug. 2018.
		
		\bibitem{CMI} A. Alkhateeb, O. El Ayach, G. Leus, and R. W. Heath,
		``Channel
		estimation and hybrid precoding for millimeter wave cellular systems,"
		\emph{IEEE J. Sel. Topics Signal Process.}, vol. 8, no. 5, pp. 831--846, Oct. 2018.
		
		\bibitem{CMII} R. M{\'e}ndez-Rial, C. Rusu, N. Gonz{\'a}lez-Prelcic, A. Alkhateeb, and
		R. W. Heath,
		``Hybrid MIMO architectures for millimeter wave communications:
		Phase shifters or switches?"
		\emph{IEEE Access}, vol. 4, pp. 247--267, Jan. 2016.
		
		\bibitem{PrecodingI} O. El Ayach, S. Rajagopal, S. Abu-Surra, Z. Pi, and R. W. Heath,
		``Spatially sparse precoding in millimeter wave MIMO systems,"
		\emph{IEEE Trans. Wireless Commun.}, vol. 13, no. 3, pp. 1499-1513, Mar. 2014.
		
		\bibitem{PrecodingII} X. Yu, J.-C. Shen, J. Zhang, and K. B. Letaief,
		``Alternating minimization
		algorithms for hybrid precoding in millimeter wave MIMO systems,"
		\emph{IEEE J. Sel. Topics Signal Process.}, vol. 10, no. 3, pp. 485-500, Apr. 2016.
		
		\bibitem{CMOMP} J. Lee, G.-T. Gil, and Y. H. Lee,
		``Channel estimation via orthogonal
		matching pursuit for hybrid MIMO systems in millimeter wave communications,"
		\emph{IEEE Trans. Commun.}, vol. 64, no. 6, pp. 2370-2386, June 2016.
		
		\bibitem{JSDM} A. Adhikary, J. Nam, J. Y. Ahn, and G. Caire,
		``Joint spatial division and
		multiplexing--The large-scale array regime,"
		\emph{IEEE Trans. Inf. Theory}, vol. 59, no. 10, pp. 6441–6463, Oct. 2013.
		
		
		\bibitem{Joint} Z. Li, S. Han, S. Sangodoyin, R. Wang, and A. F. Molisch,
		``Joint optimization of hybrid beamforming for multi-user massive MIMO downlink,"
		\emph{IEEE Trans. Wireless Commun.}, vol. 17, no. 6, pp. 3600--3614, June 2018.
		
		
		
		
		
		\bibitem{CS} S. Park and R. W. Heath,
		``Spatial channel covariance estimation for the hybrid MIMO architecture: A compressive sensing based approach,"
		\emph{IEEE Trans. Wireless Commun.}, vol. 17, no. 12, pp. 8047-8062, Dec. 2018.
		
		\bibitem{Channel estimation for TDD/FDD} H. Xie, F. Gao, S. Jin, J. Fang, and Y. C. Liang,
		``Channel estimation for TDD/FDD massive MIMO systems with channel covariance computing,"
		\emph{IEEE Trans. Wireless Commun.}, vol. 17, no. 6, pp. 4206--4218, June 2018.
		
		\bibitem{Kronecker Covariance Sketching} Y. Chi, 
		``Kronecker covariance sketching for spatial-temporal data,"
		\emph{Signal Processing Conference (EUSIPCO)}, Aug. 2016.
		
		
		\bibitem{HeroI} T. Tsiligkaridis and A. O. Hero,
		``Covariance estimation in high dimensions
		via kronecker product expansions,"
		\emph{IEEE Trans. Signal Process.}, vol. 61, no. 21, pp. 5347-5360, Nov. 2013.
		
		\bibitem{HeroII}K. Greenewald, T. Tsiligkaridis, and A. O. Hero,
		``Kronecker Sum Decompositions of Space-Time Data,"
		\emph{Proc. IEEE 5th Int.
			Workshop Comput. Adv. Multi-Sensor Adapt.
			Process}, pp. 65-68, 2013.
		
		
		
		
		\bibitem{mmWave Channel} M. Akdeniz, Y. Liu, M. Samimi, S. Sun, S. Rangan, T. Rappaport, and
		E. Erkip,
		``Millimeter wave channel modeling and cellular capacity evaluation,"
		\emph{IEEE J. Sel. Areas Commun}, vol. 32, no. 6, pp. 1164--1179, June 2014.
		
		
		\bibitem{KP}C. F. Van Loan,
		``The ubiquitous Kronecker product,"
		\emph{J. Comput. Appl. Math}, vol. 123, no. 1-2, pp. 85--100, Nov. 2000.
		
		
		\bibitem{MCRui} R. Hu, J. Tong, J. Xi, Q. Guo, and Y. Yu, 
		``Matrix completion-based channel estimation for mmWave communication systems with array-inherent impairments,"
		\emph{IEEE Access}, vol. 6, pp. 62915--62931, Oct. 2018.
		
		\bibitem{KPExp}G. Sun, Z. He, J. Tong, and X. Zhang, 
		``Knowledge-aided covariance matrix estimation via Kronecker product expansions for airborne STAP,"
		\emph{IEEE Geosci. Remote Sens. Lett.}, vol. 15, no. 4, pp. 527--531, Apr. 2018.
		
		
		\bibitem{GCG-Alt} A. Wei. Yu, W. Ma, Y. Yu, J. Carbonell, and S. Sra,
		``Efficient structured matrix rank minimization,"
		\emph{Proc. Advances in Neural Inform. Process. Syst.}, Dec. 2014.	   
		
		
		\bibitem{IMC} K. Zhong, P. Jain, and I. S. Dhillon,
		``Efficient matrix sensing using rank-1 gaussian measurements,"
		\emph{ALT 2015}, Oct. 2015.    
		
	\end{thebibliography}
\end{document}